\documentclass[prd,10pt,aps,showpacs,superscriptaddress,nofootinbib,notitlepage,floatfix,amssymb,amsmath]{revtex4-1}
\usepackage{slashed}
\usepackage{epsfig}

\newcommand{\beqn}{\begin{eqnarray}}
\newcommand{\eeqn}{\end{eqnarray}}
\newcommand{\eq}[1]{(\ref{#1})}

\newcommand{\Z}{{\mathbb Z}}

\newcommand{\bs}{\boldsymbol}

\begin{document}

\title{Chiral anomaly in Dirac semimetals due to dislocations}

\author{M.~N.~Chernodub}
\email{maxim.chernodub@lmpt.univ-tours.fr}
\affiliation{Laboratoire de Math\'ematiques et Physique Th\'eorique, Universit\'e de Tours, 37200, France}
\affiliation{Far Eastern Federal University,  School of Biomedicine, 690950 Vladivostok, Russia}

\author{M.A. Zubkov\footnote{On leave of absence from Moscow Institute of Physics and Technology, 9, Institutskii per., Dolgoprudny, Moscow Region, 141700, Russia}}
\email{zubkov@itep.ru}
\affiliation{LE STUDIUM, Loire Valley Institute for Advanced Studies, Tours and Orleans, 45000, France}
\affiliation{Laboratoire de Math\'ematiques et Physique Th\'eorique, Universit\'e de Tours, 37200, France}
\affiliation{National Research Nuclear University MEPhI (Moscow Engineering
Physics Institute), Kashirskoe highway 31, 115409 Moscow, Russia}
\affiliation{Institute for Theoretical and Experimental Physics, B. Cheremushkinskaya 25, Moscow, 117259, Russia}

\begin{abstract}
The dislocation in Dirac semimetal carries an emergent magnetic flux parallel to the dislocation axis. We show that due to the emergent magnetic field the dislocation accommodates a single fermion massless mode of the corresponding low-energy one-particle Hamiltonian. The mode is propagating along the dislocation with its spin directed parallel to the dislocation axis. In agreement with the chiral anomaly observed in Dirac semimetals, an external electric field results in the spectral flow of the one-particle Hamiltonian, in pumping of the fermionic quasiparticles out from vacuum, and in creating a nonzero axial (chiral) charge in the vicinity of the dislocation.
\end{abstract}


\date{\today}

\maketitle

\section{Introduction and Motivation}

The Dirac semimetals are novel materials that have been discovered recently (Na$_3$Bi and Cd$_3$As$_2$ \cite{semimetal_discovery,semimetal_discovery2,semimetal_discovery3}). A possible appearance of Dirac semimetals in the other systems  (for example, ZrTe$_5$ \cite{ZrTe5,ZrTe5:2}, and Bi$_2$Se$_3$ \cite{Bi2Se3}) was also discussed.  In Dirac semimetals the fermionic quasiparticles propagate according to the low energy action that has {an} emergent relativistic symmetry. Both in Na$_3$Bi and Cd$_3$As$_2$ there exist two Fermi points $\pm {\bs  K}^{(0)}$. At each Fermi point the pair of left-handed and right-handed fermions appears. The Dirac semimetals represent an arena for the observation of various effects specific for the high energy physics. In particular, the effects of chiral anomaly play an important role in physics of these materials \cite{ref:diffusion,ref:transport,semimetal_discovery3,ref:semimetal:2,ref:semimetal:3,ref:semimetal:4}.

In the Weyl semimetals, which were also discovered recently (in particular, TaAs \cite{WeylSemimetalDiscovery}) one of the two Fermi points hosts a right-handed Weyl fermion while another Fermi point hosts a left-handed Weyl fermion. Various relativistic effects were discussed in Weyl and Dirac semimetals already before their experimental discovery\cite{semimetal_effects,semimetal_effects2,semimetal_effects3,semimetal_effects4,semimetal_effects5,semimetal_effects6,semimetal_effects7,semimetal_effects8,semimetal_effects9,semimetal_effects10,semimetal_effects11,semimetal_effects12,semimetal_effects13}.

In  \cite{ZrTe5} the experimental observation of the chiral anomaly  in ZrTe$_5$ was reported as measured through their contributions to the conductance of the sample.  It has been shown, that in the presence of parallel external magnetic field and external electric field the chiral anomaly leads to the appearance of nonzero chiral density and, correspondingly, a nonzero chiral chemical potential. This work was followed be a number of papers, where the experimental detection of chiral anomaly was reported in different Dirac and Weyl semimetals (see \cite{ChiralAnomalySemimetal} and references therein).

\begin{figure}[!thb]
\begin{center}
\includegraphics[scale=0.53,clip=true]{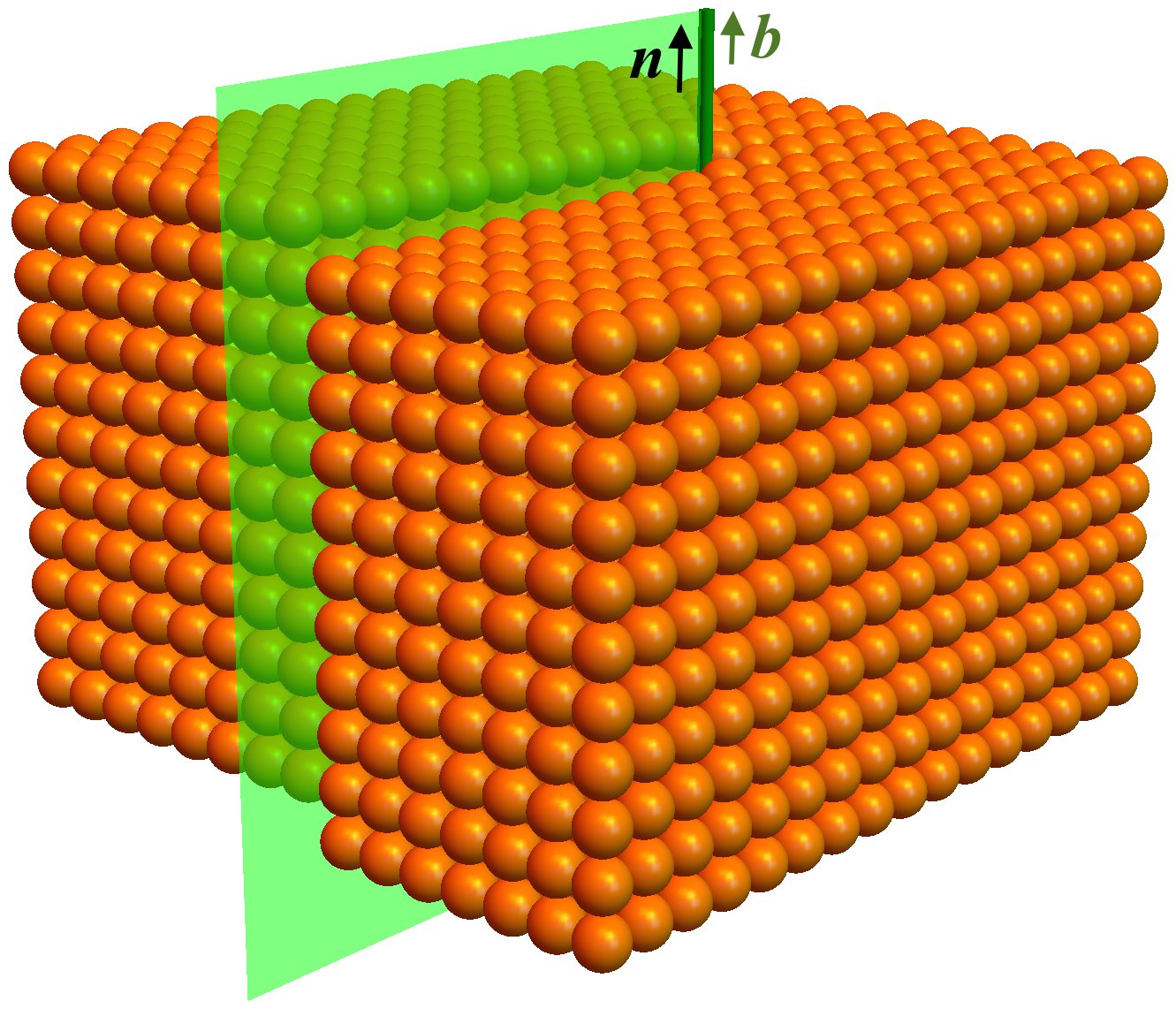}
\end{center}
\caption{Illustration of the screw dislocation of the atomic lattice with the Burgers vector ${\bs b}$ parallel to the axis ${\bs n}$ of the dislocation (the green line). The semitransparent plane points out to the region where the atomic planes experience a shift.}
\label{fig:screw}
\end{figure}

\begin{figure}[!thb]
\begin{center}
\includegraphics[scale=0.57,clip=true]{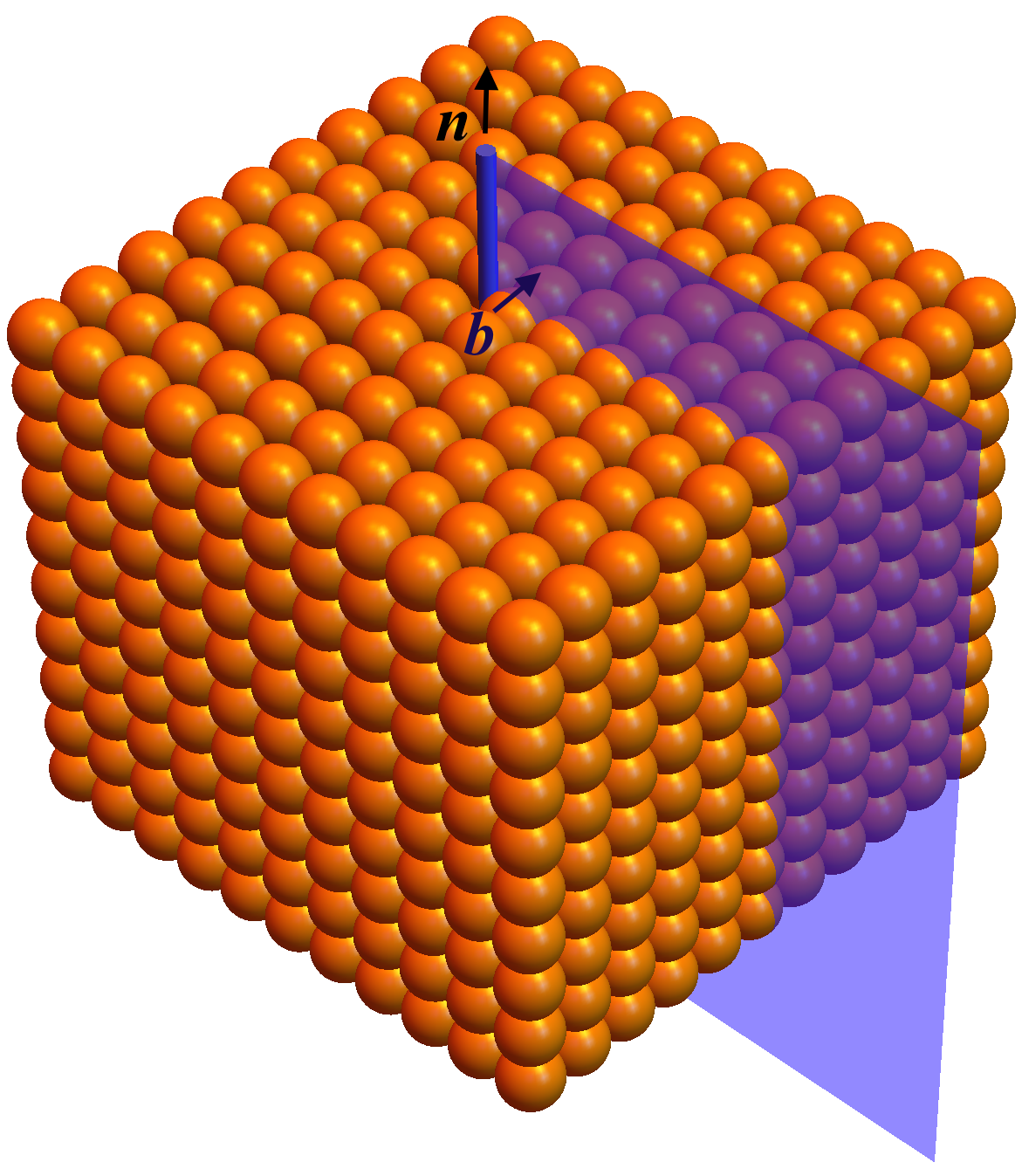}
\end{center}
\caption{Illustration of the edge dislocation with the Burgers vector ${\bs b}$ perpendicular to the dislocation axis ${\bs n}$ (the blue line). The semitransparent plane shows the extra half-plane of ions introduced in the crystal.}
\label{fig:edge}
\end{figure}

Similarly to graphene \cite{vozmediano2,vozmediano3,vozmediano4,vozmediano5,vozmediano6,VZ2013,VolovikZubkov2014,Volovik:2014kja}
the fermionic quasiparticles in Dirac and Weyl semimetals experience emergent gauge field and emergent gravity in the presence of elastic deformations of the atomic lattices (see, for example, \cite{Volovik2003,Parrikar2014,Vozmediano,Z2015} and references therein). In this paper we will concentrate on dislocations in the crystalline order of the atomic lattice, which are particularly interesting cases of the elastic deformations of the ion crystal lattice~\cite{ref:LL,ref:dislocations}. The dislocation is a line-like defect characterized by the Burgers vector $\bs b$ which determines the physical displacement of the atomic lattices along the dislocation. The vector ${\bs b}$ is a global characteristic of the dislocation because it is a constant quantity over the entire length of the dislocation. In rough terms, one may imagine the dislocation as a vortex which possesses a fixed ``vorticity'' given by the Burgers vector $\bs b$. The extreme examples of the dislocations are the screw dislocation (shown in Fig.~\ref{fig:screw}) and the edge dislocation (illustrated in Fig.~\ref{fig:edge}) for which the corresponding Burgers vectors are parallel and, respectively, perpendicular to dislocations' axes ${\bs n}$. There are other types of the dislocations lying in between these two extreme cases.

In \cite{Z2015} the effect of the dislocation on the geometry experienced by fermionic quasiparticles in Dirac semimetals was considered for the first time. Aharonov-Bohm effect and Stodolsky effect (the
latter effect describes a correction to the Aharonov-Bohm effect due to torsion) were investigated for the scattering of the quasiparticles on dislocations. Besides, basing on an obvious analogy with the results of \cite{ZrTe5} it was proposed, that the dislocation (that carries an emergent magnetic flux) becomes the source of chiral anomaly and chiral magnetic effect. This occurs because the dislocation carries emergent magnetic field. Therefore, it was argued, that the chiral anomaly and chiral magnetic effect occur without any external magnetic field.   According to \cite{Z2015} the contribution of topology to magnetic flux $\Phi$ is equal to the scalar product ${\bs  K}^{(0)}{\bs  b}$, where $\bs  b$ is the Burgers vector. There may also appear the contribution to the flux $\Phi$ proportional to the tensor of elastic deformations caused by the dislocation with the coefficients of proportionality that are analogous to the Gruneisen parameter of graphene.
The emergent magnetic flux is associated with emergent magnetic field.
\begin{equation}
{H}^i(x) = \Phi\, \int d y^i \delta^{(3)}(x - y),\label{Hd}
\end{equation}
where the integral is taken along the dislocation. The appearance of the delta-function in Eq.~\eq{Hd} in the low-temperature theory corresponds to the fact that the emergent magnetic flux is localized within the dislocation core of radius $\xi \sim |{\bf b}|$, where ${\bf b}$ is the Burgers vector of the dislocation. In \cite{Z2015} the simple model of the dislocation was used, in which it is represented as a tube of size $\xi$ with the emergent magnetic field inside it.

The further examination of the mentioned above problem has led us to the conclusion, that the naive application of the pattern of chiral anomaly discussed in \cite{ZrTe5} to the case, when the magnetic field is emergent and is caused by dislocations, has certain restrictions. Strictly speaking, the mentioned above model of the fermionic excitations and chiral anomaly within the dislocation may be applied to the investigation of real materials only if the emergent magnetic flux of the low energy field theory is distributed within the area of size $\xi$ essentially larger, than the interatomic distance $a$ while the emergent magnetic flux of the dislocation is essentially larger, than $2\pi$. In this situation we are formally able to use the low energy field theory for the description of fermionic excitations inside the dislocation core. This occurs for the strong dislocations with sufficiently large values of the Burgers vector $\bs  b$, when the crystal lattice is distorted considerably (or in the case, when many parallel dislocations with small values of the Burgers vectors are located close to each other). In this case the dislocation core size $\xi \sim |{\bf b}| \gg a$ is much larger than the interatomic distance $a$.

In the present paper we consider the opposite situation, when in Dirac semimetals the values of Burgers vector are relatively small, so that the magnetic flux at the dislocation is smaller than $2\pi$ or around $2\pi$. In this situation the crystal structure is not violated strongly, so that the dislocation core size is, presumably, of the order of the interatomic distance $\xi \sim a$. The low energy theory is developed for the states with the typical values of momenta much smaller, that $1/a$. Therefore, in this case the states localized within the dislocation core cannot be described by the field theory. In order to describe such states the microscopic theory is to be applied.

It appears, that the anomaly in the right-  and left-handed quasiparticle currents is given by
\begin{equation}
\langle \partial_\mu j^\mu_{R,L} \rangle = \pm \frac{1}{4\pi^2} {\bs  E} {\bs  B}
\quad \mbox{or}\quad
\langle \partial_\mu j^\mu_5 \rangle =
 \frac{1}{2\pi^2} {\bs  E} {\bs  B}\,,
 \label{CA0}
\end{equation}
where the upper and lower signs in the first equation correspond to the right-handed and {the} left-handed quasiparticles, respectively. The important feature of  Eq.~\eq{CA0} is that the effective magnetic field ${\bs  B}$ -- contributing to the anomaly~\eq{CA0} differs from the emergent magnetic field at the dislocation ${\bs H}$ as given in Eq.~(\ref{Hd}).

The basic reason for the difference between the emergent magnetic field ${\bs H}$ and the effective magnetic field ${\bs B}$ is that the emergent magnetic field ${\bs H}$ of the dislocation has a {small} (of the order of unity or even smaller) magnetic flux $\Phi$. In this case the contribution to both mentioned effects is given by {the} single fermionic mode (related to the zero mode of the one-particle Hamiltonian) propagating along the dislocation rather {than} by a large ensemble of the lowest Landau modes with a huge degeneracy factor. To be more precise, the effective magnetic field $\bs  B$ is expressed through the probability density corresponding to the zero mode of the one-particle Hamiltonian in the background of the emergent magnetic field $\bs H$ due to the dislocation. The appearance of the propagating (zero) mode at the dislocation is a natural effect, which is known to exist in topological insulators with lattice dislocations~\cite{ref:topins}.

The contribution of the individual zero mode to the chiral anomaly~\eq{CA0} can be described with the help of the effective magnetic field ${\bs B}$, which carries {a unit elementary magnetic flux contrary to the original emergent magnetic field ${\bs H}$, which may have an arbitrary (but still small) value of the total flux $\Phi$}. The effective field $\bs  B$ is localized in the wide area of linear size $\xi_0$, where $1/\xi_0$ is the infrared cutoff of the considered field theoretical low energy approximation (below we argue that $\xi_0$ may be identified with the mean free path of the quasiparticles. For example, in Cd$_3$As$_2$ the mean free path is $\xi_0 \sim 200 \,\mu$m \cite{ref:transport}).

In this paper we demonstrate that for {the} straight screw dislocation directed, for example, along the symmetry axis of the crystal the emergent magnetic flux associated with the emergent field ${\bs H}$ is given by
\begin{equation}
{\Phi} = \bigl({\bs  K}^{(0)} \cdot {\bs  b}\bigr) + \frac{\beta}{2 a} ({\bs n} \cdot {\bs b})\,,
\label{eq:Phi:intro}
\end{equation}
where the first term is of the topological origin~\cite{Z2015} while the second term is not topological (here $\beta$ is an analogue of the Gruneisen parameter of graphene \cite{VZ2013}).  The magnetic field associated with the flux~\eq{eq:Phi:intro} is localized within the dislocation core of a typical size $\xi \sim a$, where $a\sim 1\, n$m is a typical interatomic distance. In Eq.~\eq{eq:Phi:intro} the vector ${\bs  K}^{(0)}$ encodes positions of the Fermi points ${\bs k} = \pm {\bs  K}^{(0)}$ in {momentum space,} and ${\bs n}$ is the direction of the dislocation axis. For {the} straight screw dislocation {vectors} ${\bs  K}^{(0)} $, $ {\bs  b}$ and ${\bs n}$ in Eq.~\eq{eq:Phi:intro} are parallel to each other.

We will discuss effects, which appear due to the interplay between quantum anomaly and dislocations in the crystal structure of Dirac semimetals.  A fermion excitation is affected by the dislocation, in particular, via the mentioned above intrinsic magnetic field, which is localized in a spacial vicinity of the dislocation and {is} directed along the axis of the dislocation.
In principle, the emergent magnetic fields corresponding to different Weyl fermions (that belong to different Fermi points and/or have different chiralities) differ from each other. However, there exists an approximation, in which those emergent fields $\bs  H$ have the same absolute values, but opposite directions for the two Fermi points $\pm {\bs  K}^{(0)}$. If this approximation is not violated strongly (which is the general case) the signs of the emergent magnetic fluxes experienced by the quasiparticles living near to the Fermi points $\pm {\bs  K}^{(0)}$ are opposite. In the Dirac semimetal both right- and left-handed fermion excitations are present in each (of the two) Dirac cone, therefore in this case we have a standard effective magnetic field ${\bs B}({\bs x})$ acting on the right- and left-handed fermions at one Dirac cone and the magnetic field $-{\bs B}({\bs x})$ acting on the right- and left-handed fermions at another cone. These fields enter expression for the anomaly Eq. (\ref{CA0}).

If now one applies an external static electric field ${\bs E}$ along the axis of the dislocation, then the quantum anomaly will generate the chiral charge at a rate proportional to the scalar product ${\bs E}{\bs B}$. The generated chiral charge will dissipate, both due to {the} chiral-changing processes inside the region of size $\xi_0$ around the dislocation and due to the spatial diffusion of the chiral charge.  Next we notice that the equilibrium distribution of the chiral charge -- which can effectively be described by a spatially nonconstant but otherwise static chiral chemical potential $\mu_5$ -- is subjected to the intrinsic magnetic field of the dislocation itself. The chemical potential $\mu_5$ is distributed around the dislocation with the characteristic length $L_V$ (for example, in Cd$_3$As$_2$ this length is of the order of $L_V \sim 2\, \mu$m.) The chirally imbalanced matter in the presence of magnetic field generates dissipationless electric current directed along the dislocation and concentrated in the spatial vicinity around it. Therefore, the intrinsic magnetic field of the dislocation would lead to a spatially-dependent (negative) magnetoresistance around the dislocation. Similar arguments were used in Ref.~\cite{ZrTe5} to experimentally investigate effects of the chiral anomaly in ZrTe${}_5$ in the presence of external magnetic field.

The paper is organized as follows. In Sect. \ref{SectDWS} we recall briefly general theory of quasi-relativistic fermions in Dirac semimetals in the presence of elastic deformations which {leads both to the emergent gauge field and to the emergent gravity} (the latter is described by an emergent vielbein~\cite{VZ2014NPB}). In Section~\ref{SectEMD} we discuss these effects focusing on dislocations, partially following Ref.~\cite{Z2015}. In Sect.~\ref{SectZMD} we consider {the} zero modes of the one-particle Hamiltonian and demonstrate, that there always exists a single mode with {the} definite spin directed along the emergent magnetic flux, which is localized in a wide area around the dislocation. In Sect.~\ref{SectCAWDS} we show that the spectral flow along the branch of spectrum (that crosses zero at the mentioned zero mode) gives rise to the anomalies in quasiparticle currents: in a Dirac semimetal the chiral anomaly appears. For the sake of simplicity, these results are discussed first for {the strait dislocation,} is directed along the symmetry axis $z$ of the crystal {that} coincides with the direction of the Fermi point ${\bs  K}^{(0)}$ in {momentum space}. We extend our results to the case of strait dislocations with arbitrary direction in Sect.~\ref{SectAnis}. Then in Sect.~\ref{SectCDS} we discuss the generation of the chiral charge via the chiral anomaly~\eq{CA0} due to {the} interplay between an external electric field and {the} internal magnetic field of the dislocation. The last section is devoted to {discussions} and {to} our conclusions.

\section{Relativistic fermions in Dirac Semimetal}
\label{SectDWS}

The Dirac semimetal possesses two cones, each of which hosts one right-handed and one left-handed Weyl fermion. In the presence of elastic deformations caused by the dislocation the action for a right-handed and left-handed Weyl fermions near a given Fermi point are, respectively, as follows~\cite{Z2015}:
\beqn
S_R & {=} & \frac{1}{2} \!\int d^4x |{\bs  e}| \! \left[\bar{\Psi} i e_b^\mu(x) {\sigma}^b {\cal D}_\mu  \Psi - [{\cal D}_\mu\bar{\Psi}] i e_b^\mu(x) {\sigma}^b  \Psi \right]\!, \quad\
\label{SHe_3sW2} \\
S_L  & {=} & \frac{1}{2} \!\int d^4x |{\bs  e}| \! \left[\bar{\Psi} i e_b^\mu(x) \bar{\sigma}^b {\cal D}_\mu \Psi - [{\cal D}_\mu\bar{\Psi}] i e_b^\mu(x) \bar{\sigma}^b  \Psi \right]\!,
\label{SHe_3sW6}
\eeqn
where
\begin{equation}
i{\cal D}_\mu = i\nabla_\mu + A_\mu(x)
\end{equation}
is the covariant derivative corresponding to the emergent $U(1)$ gauge field $A_\mu$, $\sigma^0 = \bar{\sigma}^0 = 1$, and $\bar{\sigma}^a = - \sigma^a$ with $a=1,2,3$ are the Pauli matrices. The currents of the right- and left-handed quasiparticles are, respectively, as follows:
\beqn
J^\mu_R & = & \bar{\Psi} i e_b^\mu(x) {\sigma}^b  \Psi\,, \label{JR} \\
J^\mu_L & = & \bar{\Psi} i e_b^\mu(x) \bar{\sigma}^b  \Psi\,. \label{JL}
\eeqn
Throughout this paper the internal $SO(3,1)$ indices are denoted by Latin letters $a,b,c,...$ from the beginning of the alphabet while the space-time indices are denoted by Greek letters or by Latin letters $i,j,k,...$ from the middle of the alphabet.

The vierbein field $e_a^\mu = e_a^\mu(x)$ is a $4 \times 4$ matrix which carries all essential information about anisotropy and the elastic deformations (caused, for example, by a dislocation) of the ion lattice of the Dirac crystal. It is convenient to introduce the inverse of the inverse vierbein field, $e_\mu^a = e_\mu^a(x)$, defined, naturally, as follows:
\beqn
{e}^\mu_a (x) {e}^a_\nu (x) = \delta^\mu_\nu\,.
\eeqn
In our paper we always assume that the deformations are small so that the determinant of the vierbein field
\beqn
|{\bs  e}| \equiv \det ({\bs  e}^a_\mu)\,,
\label{eq:det:e}
\eeqn
never vanishes.

In the absence of elastic deformations the fields entering the actions~\eq{SHe_3sW2} and \eq{SHe_3sW6} are simplified. In this case the emergent gauge field $A_\mu$ vanishes.

In the absence of elastic deformations the vierbein can be chosen in a diagonal form,
\begin{equation}
e^{(0),\mu}_a = \left(
\begin{array}{cccc}
v^{-1}_F & 0 & 0 & 0 \\
0 & \nu^{-1/3} & 0 & 0 \\
0 & 0              & \nu^{-1/3}& 0 \\
0 & 0 & 0      & \nu^{2/3}
\end{array}
\right),
\label{Connection00}
\end{equation}
where the parameter $\nu \neq 1$ reflects the fact that the experimentally studied Dirac semimetals are anisotropic materials~\cite{semimetal_discovery,semimetal_discovery2,semimetal_discovery3}. It is also convenient to introduce the spatial component of the undeformed vierbein~\eq{Connection00}:
\beqn
e^{(0),i}_a
\equiv \hat{f}^i_a
= \left(
\begin{array}{ccc}
\nu^{-1/3} & 0 & 0 \\
0              & \nu^{-1/3}& 0 \\
0 & 0      & \nu^{2/3}
\end{array}
\right),
\label{eq:hat:f}
\eeqn
with $i,a = 1,2,3$. The quantity $v^{i}_F \equiv v_F \hat{f}^i_i$ with fixed $i=1,2,3$ has a meaning of the anisotropic Fermi velocity in $i$-th direction. The determinant~\eq{eq:det:e} in the undeformed case is $|{\bs  e}^{(0)}| = v_F$.

The low-energy effective field theory~\eq{SHe_3sW2}, \eq{SHe_3sW6} has the natural ultraviolet cutoff $\Lambda_{UV} \sim |{\bs  K}^{(0)}|$ associated with the positions of the Dirac cones in the momentum space. In order to determine a natural infrared cutoff we notice that in our field-theoretical approximation the massless quasiparticles do not interact with each other since the effective actions~\eq{SHe_3sW2} and \eq{SHe_3sW6} contain only bilinear fermionic terms while the gauge field $A_\mu$ is  a classical non-propagating field. Therefore,  the natural infrared cutoff for our approach is $\Lambda_{IR} = 1/\xi_0$, where the length $\xi_0$  may be identified with the mean free path of the massless quasiparticles. Indeed at the distances of the order of the mean free path $\xi_0$ we cannot neglect interactions between the quasiparticles and their scattering on the defects of the atomic lattice which, in general, cannot be captured by Eqs.~\eq{SHe_3sW2} and \eq{SHe_3sW6}.

As an example, we mention that for the Dirac material Cd$_3$As$_2$ the mean free path $\xi_0$ was estimated in Ref.~\cite{ref:transport} to be of the order of $200 \,\mu$m. In the above  formulation of the low-energy theory, the Dirac point corresponds to zero energy. In real situation the crystals of semimetal may have nonzero Fermi energy at the level crossing points. In particular, in \cite{EF} the values of Fermi energy of the order of $10$ meV were reported for Na$_3$Bi. In the following applications we assume, that in the real systems the value of Fermi energy may be neglected, or that the sample is doped in such a way, that  the doping-induced chemical potential shifts the level crossing to the vanishing energy.

In the upcoming sections for simplicity we restrict ourselves to the case, when the dislocation is an infinite straight line directed along the symmetry axis $z$ of the crystal, which coincides with the direction of the Dirac point ${\bs  K}^{(0)}$ in {momentum space}. We will return to {the} more general case of an arbitrarily aligned straight dislocation in Sect. \ref{SectAnis}.

Now let us consider the case when the atomic lattice of a Dirac semimetal is elastically deformed. The deformation is described by the displacement vector $u^i$ which gives the displacements of the ions with respect to their positions with respect to the unperturbed semimetal. In the approximation of isotropic elasticity for a straight dislocation directed around the $z \equiv x_3$ axis the displacement vector $u^a$  is given by:
\begin{equation}
u^a = -  \theta \frac{b^a}{2\pi}  + u^a_{\rm cont},
\label{using}
\end{equation}
where $\theta$ is the polar angle in the plane orthogonal to the dislocation and $b^a$ is the Burgers vector. The first term in the right hand side of Eq.~\eq{using} is discontinuous vector function as it has a jump by $b^a$ at $\theta = 0$. The second, continuous part of displacement is given by~\cite{ref:LL:problems}
\beqn
u^k_{\rm cont}({\bs x}_\perp) {=}  -\frac{b^l}{4\pi} \frac{1-2\sigma}{1-\sigma} \left[{\epsilon}^{3kl} {\rm log} \frac{|{\bs x}_\perp|e^{\gamma}}{2 R}
{+} \frac{{\epsilon}^{3il} \hat{x}_\perp^i \hat{x}^k_\perp}{1-2\sigma}\right] \qquad
\label{ucontsol1}
\eeqn
where ${\bs x}_\perp = (x_1,x_2)$ are the transverse coordinates in the laboratory reference frame, $\hat{x}_\perp^i$ are respective unit angles in the transverse plane and $\sigma$ is the Poisson ratio which is defined as the negative ratio of transverse to axial strain of the atomic crystal. Throughout this paper we shall work in the laboratory reference frame in which the positions of ions are their real 3d coordinates.

Notice, that for the screw dislocation when the Burgers vector directed along the dislocation axis, ${\bs b} = (0,0,b_z)$, the continuous part of the displacement vector vanishes, $u^k_{\rm cont} = 0$. It is worth mentioning, that while the values of $u_{\rm cont}^k$ may be large, its derivatives are small for sufficiently small $b$ because after the differentiation the expression in Eq.~(\ref{ucontsol1}) tends to zero at $|{\bs x_\perp}|\rightarrow \infty$.

In the presence of elastic deformations, in principle, the emergent vielbeins (as well as the emergent gauge fields) may differ for the left-handed and the right-handed fermions incident at the given Dirac point.

Let us introduce tensor of elastic deformations~\cite{ref:LL}
\beqn
u^{ij} = \partial^i u^j + \partial^j u^i\,,
\label{eq:uij}
\eeqn
where we have neglected {the} part quadratic in $u^i$ by assuming that the deformations are small. In general, the emergent vielbein around the dislocation may be expressed, up to the terms linear in displacement vector, as follows (see Ref.~\cite{Z2015} for the details of the derivation):
\begin{eqnarray}
  e_a^{i} & = & \hat{f}_a^{i}(1+ \frac{1}{3} \gamma^k_{k n j} u^{nj} )
  + \hat{f}^k_a \partial_k u^{i} - \hat{f}^n_a \gamma^i_{n jk} u^{jk}\nonumber\\  e_0^{i} & = & - \frac{1}{v_F} \gamma^i_{0 jk} u^{jk}, \quad  e^0_a = 0\nonumber\\
  e_0^0 & = & \frac{1}{v_F}(1 + \frac{1}{3} \gamma^k_{k ij} u^{ij})
 \nonumber\\
 |{\bs  e}| & = &  v_F(1 - \partial_i u^i - \frac{1}{3} \gamma^k_{k ij} u^{ij})\nonumber\\ &&  a,i, j, k,n = 1,2,3
\label{eg}
\end{eqnarray}
The emergent gauge field is given by
\begin{eqnarray}
  A_i & \approx &  -  \nabla_i({\bs  u}\cdot {\bs  K}^{(0)})) +    \frac{1}{a}\beta^{}_{ijk} u^{jk},
  \label{DiracPosition1}\\
  A_0^{} &=& \frac{1}{a}\beta^{}_{0jk} u^{jk}, \quad i,j,k=1,2,3
 \nonumber
\end{eqnarray}
The tensors $\beta$ and $\gamma$, which are the analogues to the Gruneisen parameters in graphene, may, in principle be different for the right-handed and the left-handed fermions. The analogy to graphene prompts that their values could be of the order of unity. Notice, that in graphene the emergent electric potential $A_0$ does not arise outside of the dislocation core~ \cite{VZ2013}. In the same way we assume, that in the semimetal the parameters $\beta_{0jk}$ may be neglected. The reason for this is that the combination ${\bs  K}^{(0)} + {\bs  A}$ appears as the value of momentum ${\bs  P}$, at which the one-particle Hamiltonian ${\cal H}({\bs  x},\hat{\bs  P})$ vanishes (one substitutes ${\bs  K}^{(0)} + {\bs A}$ instead of the momentum operator $\hat{\bs P}$):
\begin{equation}
{\cal H}\bigl({\bs x},{\bs K}^{(0)} + {\bs A}({\bs x})\bigr) = 0
\end{equation}
As a result we expand the Hamiltonian near the floating Fermi point ${\bs K}^{(0)} + {\bs A}({\bs x})$:
\beqn
{\cal H}({\bs x},\hat{\bs P}) & = & |{\bs e}({\bs x})| \,  e_a^k({\bs x}) \sigma^a \circ \left[\hat P_k - \bigl(K^{(0)}_k +  A_k({\bs x})\bigr)\right]  \nonumber\\
& & +  A_0({\bs x})\,,
\eeqn
where by the symbol $\circ$ we denote the symmetric product
\beqn
A\circ B = \frac{1}{2}(AB + BA).
\eeqn
The only possible source of $ A_0({\bs x})$ is the noncommutativity of momentum $\hat{\bs P}$ and coordinates. This means, that unlike $ A_k$ with $k = 1,2,3$ the emergent electric potential $ A_0$ is proportional to the derivatives of the parameters entering ${\cal H}({\bs x},\hat{\bs P})$.  The field $ A_k$ with $k = 1,2,3$ is proportional to $1/a$ times the combination of the dimensionless parameters while $ A_0$ is proportional to their derivatives but it does not contain the factor $1/a$. For slow varying elastic deformations this means that $ A_0$ may be neglected. This consideration does not work, however, within the dislocation core, where physics is much more complicated. The influence of this unknown physics on the quasiparticles with small values of momenta (described by the action of the form of Eqs. (\ref{SHe_3sW2}), (\ref{SHe_3sW6})) may be taken into account through the same emergent fields $ A_{\mu}, \mu = 0,1,2,3$ and $ e^k_a$, which become strong within the dislocation core. The component of $ A_0$ of emergent electromagnetic field is not forbidden by any symmetry. Therefore, it appears and gives rise to emergent electric potential (either attractive or repulsive) within the dislocation core.

Notice, that the simple model of Weyl semimetal with cubic symmetry has been considered in \cite{Vozmediano}. {The} Dirac semimetal (with cubic symmetry) may, in principle, be described by the two copies of the model of~\cite{Vozmediano}.

\section{Emergent magnetic flux carried by the dislocation}

\label{SectEMD}

In order to calculate the emergent magnetic field we should use integral equation
\begin{equation}
\frac{1}{2}{\epsilon}_{ijk}\int_{\cal S} {H}^i dx^j \wedge dx^k \equiv \int_{\partial {\cal S}}  A_k dx^k\,, \label{AdefI}
\end{equation}
where the integration goes over a surface in the transverse plane which includes the position of the dislocation. For the considered solution of elasticity equations~\eq{DiracPosition1}  we represent the right hand side of this expression as follows
\begin{equation}
\int_{\partial {\cal S}}  A_k dx^k =   b^i  K^{(0)}_i +  \frac{1}{a}\beta^{}_{ijk} \int_{\partial {\cal S}} u^{jk} dx^i
\label{AdefI2}
\end{equation}
The first term in this expression gives the following singular contribution to magnetic field:
\begin{equation}
{H}^k_{{\rm sing}}({\bs x})
\approx  b^i  K^{(0)}_i \int_{l^0}  d y^k(s) \delta^{(3)} (x-y(s)),\label{Hsing}
\end{equation}
where the integration over $y$ goes along the dislocation axis $l^0$.

One can check that the solutions of elasticity equations give $u^{jk} \sim 1/r$ at $r\rightarrow \infty$. Therefore, the integral along the circle ${\cal C}_r \equiv \partial {\cal S}$ at $r\rightarrow \infty$ (with the dislocation at its center) in the second term of Eq. (\ref{AdefI2}) gives finite contribution to the normalized total flux of the singular gauge field ${\bs H}_{\mathrm{sing}}$:
\begin{equation}
\hat{\Phi}(r) = \frac{\Phi(r)}{\Phi_0} =  \frac{1}{2\pi}\int_{{\cal C}_r} A_k dx^k\,,
\label{eq:flux}
\end{equation}
where
\beqn
\Phi_0 = 2 \pi
\label{eq:Phi:0}
\eeqn
is the elementary flux (in out units the electric charge is unity $e=1$). At the same time the function $\Hat{\Phi}(\infty) - \Hat{\Phi}(r)$ takes its maximum at $r=0$ and decreases fast out of the core of the dislocation.

In the considered crystals there exist several exceptional vectors ${\bs G}_i$ ($i = 0,1,2,...$), which generate the symmetry of Brillouin zone, i.e. momenta ${\bs k}$ and ${\bs k} + {\bs G}_i$ are equivalent.
The unperturbed Fermi point is directed along ${\bs G}_0$ and is also defined up to the transformations ${\bs K}^{(0)} \rightarrow {\bs K}^{(0)} + {\bs G}_i$. This corresponds to the change of the magnetic flux by
\beqn
\Delta \Hat{\Phi}={\bs b}\cdot{\bs G}_i = 2\pi N\,,
\qquad
N \in \Z\,.
\label{eq:Burgers:condition}
\eeqn
Such a change of the magnetic flux is unobservable for Weyl fermions
and Eq.~\eq{eq:Burgers:condition} is thus posing certain restrictions on the choice of the Burgers vectors.
For example, for the layered hexagonal structure of Na and Bi atoms in the compound Na$_3$Bi  we have
\begin{eqnarray}
&&{\bs G}_1 = \frac{4\pi}{3a} \hat{\bs x}, \quad {\bs G}_2 = \frac{4\pi}{3a}\Bigl(\frac{1}{2}\hat{\bs x} + \frac{\sqrt{3}}{2}\hat{\bs y}\Bigr), \nonumber\\ &&  {\bs G}_3 = \frac{4\pi}{3a}\Bigl(\frac{1}{2}\hat{\bs x} - \frac{\sqrt{3}}{2}\hat{\bs y}\Bigr), \quad {\bs G}_0 = \zeta \frac{4\pi}{3a}\hat{\bs z}.
\end{eqnarray}
Here $a$ is the interatom distance within each layer in the plane orthogonal to ${\bs G}_0 \parallel {\bs K}^{(0)}$ and the material parameter $\zeta$ determines the interlayer distance $a_z = \frac{3a}{2\zeta}$. Due to the hexagonal (honeycomb) structure of the Na$_3$Bi layers in $xy$ plane, we may construct the Burgers vectors similarly to the case of graphene~\cite{Volovik:2014kja} which has also the hexagonal structure. Condition~\eq{eq:Burgers:condition} gives us the following general expression for the Burgers vectors:
\begin{equation}
{\bs b} = \sum_i N_i {\bs m}_i
\end{equation}
where $N_i \in \Z$. The vectors ${\bs m}_i$
\begin{eqnarray}
{\bs m}_0 & = & \frac{3a}{2\zeta}\hat{\bs z}\,,\nonumber\\
{\bs m}_1 & = & - {\bs l}_1 + {\bs l}_2 \equiv \frac{3a}{2}\hat{\bs x}+\frac{\sqrt{3}a}{2}\hat{\bs y}\,, \nonumber\\
{\bs m}_2 & = & {\bs l}_3 - {\bs l}_2 \equiv  -\sqrt{3}a \hat{\bs y}\,,\\
{\bs m}_3 & = & {\bs l}_1 - {\bs l}_3 \equiv -\frac{3a}{2}\hat{\bs x}+\frac{\sqrt{3}a}{2}\hat{\bs y}\,,\nonumber
\end{eqnarray}
are constructed from the
unit vectors ${\bs l}_1$, ${\bs l}_2$ and ${\bs l}_3$ which correspond to the nearest-neighbor Na-Bi bonds of the honeycomb lattice in the transverse planes of Na$_3$Bi:
\begin{eqnarray}
{\bs l}_1 & = & - a \, \hat{\bs x},  \nonumber\\
{\bs l}_2 & = & a\, \Bigl(\frac{1}{2}\hat{\bs x} + \frac{\sqrt{3}}{2}\hat{\bs y}\Bigr),
\label{eq:lll}\\
{\bs l}_3 & = & a\, \Bigl(\frac{1}{2}\hat{\bs x} - \frac{\sqrt{3}}{2}\hat{\bs y}\Bigr)\,.
\nonumber
\end{eqnarray}

For a screw dislocation perpendicular to the layers of Na$_3$Bi the displacement vector is given by Eq.~\eq{using}. The only nonzero components of the corresponding deformation tensor~\eq{eq:uij} are
\beqn
u^{3a} ({\bs x}_\perp) \equiv u^{a3} ({\bs x}_\perp)
= \frac{b_3\epsilon^{3ab}x^b_\perp}{4\pi {\bs x}_\perp^2}\,,
\label{eq:u3a}
\eeqn
and the emergent electromagnetic field~\eq{DiracPosition1} is given by the following expression:
\begin{eqnarray}
 A_i =  - \nabla_i({\bs u}{\bs K}) + \frac{\beta}{a} u^{3i} + \frac{\beta^{\prime}}{a} \epsilon_{3ij} u^{3j} \,, \quad\
 A_0 =  0\,, \quad  \label{ANa3Bi}
\end{eqnarray}
with some material-dependent constants $\beta$ and $\beta^{\prime}$. Notice that our expression~\eq{ANa3Bi} differs from that of Ref.~\cite{Vozmediano}. Equation~\eq{ANa3Bi} leads to the following expression for the (normalized) magnetic flux~\eq{eq:flux}
of the emergent magnetic field ${\bs H}$:
\begin{equation}
\Hat{\Phi}(\infty) = \frac{{\bs K}^{(0)}{\bs b}}{2\pi} + \frac{\beta}{4 \pi a} b_3
\label{PhiNa3Bi}
\end{equation}
For example, in Na$_3$Bi the value of ${\bs K}^{(0)} \approx 0.26 \frac{\pi}{a_z}\, \hat{z}$, where $a_z$ is the lattice spacing in $z$ direction~\cite{semimetal_discovery,ref:kzDirac}.
The value of $b_3 = N a_z$ is proportional to  $a_z$. Therefore, the topological contribution to magnetic flux of the dislocation is $\frac{{\bs K}^{(0)}{\bs b}}{2\pi} \approx \frac{0.26 \pi N}{2\pi} \approx 0.13\, N$. Following an analogy to graphene, where Gruneisen parameter $\beta \sim 2$ we may roughly estimate the second term in Eq. (\ref{PhiNa3Bi}) as $\sim 0.2\, N$. Then the emergent magnetic flux incident at the dislocation, presumably, reaches the value of $2\pi$ at $N \sim 30$.

As it was mentioned above, we may neglect the zero component of the emergent electromagnetic field $ A_0$ at large distances $r \gg a$, where the elasticity theory works. However, such a potential may be present within the dislocation core because of the essential change in the microphysics at the interatomic scales. Thus we assume the existence of either repulsive or attractive potential
\beqn
A_0({\bs x}_\perp) = v_F \nu^{-1/3}\phi({\bs x}_\perp)\,,
\eeqn
localized at the dislocation core. We neglect possible appearance of such potential far away from the dislocation axis.

\section{Fermion zero modes propagating along the dislocation}
\label{SectZMD}

Momentum of quasiparticles that are described by the action of Eq. (\ref{SHe_3sW2}) should be much smaller than $1/a$. At the same time the emergent gauge field ${\bs A}$ within the dislocation core may be as large as $\sim 1/a$. Therefore, the contribution of emergent gravity to Eq. (\ref{SHe_3sW2}) is always small compared to the contribution of the emergent gauge field. Nevertheless, for the completeness we consider this contribution in Appendix A.

In this section we show that the dislocations in a Dirac semimetal host a topologically protected massless (quasi)fermion mode, which propagates along the dislocation with the Fermi velocity. This fermionic mode corresponds to the zero mode of the transverse Hamiltonian~\eq{eq:H:R:transverse}. The appearance of the propagating mode localized in the vicinity of the dislocation is also known to emerge in topological insulators with lattice dislocations~\cite{ref:topins}.

Let us neglect the emergent gravity due to its weakness and concentrate first on the case of {the} screw dislocation, when $ A_3=0$. We may apply the gauge transformation, which brings the gauge field to the form
\begin{equation}
A_i = \epsilon_{3ij}\partial_j f({\bs x}_\bot)\,,
\label{eq:A:if}
\end{equation}
where $f$ is a function of transverse coordinates. Then the Hamiltonian~\eq{eq:HR} receives the form:
\begin{eqnarray}
{\cal H}^{(R)}  &=& v_F\nu^{2/3} \sigma^3  \hat{p}_3   + v_F \nu^{-1/3}{\cal H}^{(R)}_\bot
\end{eqnarray}
with
\begin{eqnarray}
{\cal H}^{(R)}_\bot & \approx & \phi(x_\bot) + \sum_{a=1,2}\sigma^a  \Big(\hat{p}_a + A_a({\bs x})\Bigr). \qquad
\label{eq:HR:2}
\end{eqnarray}

The zero modes of the transverse Hamiltonian~\eq{eq:HR:2} are defined as the solutions of equation
\begin{equation}
{\cal H}^{(R)}_\bot \psi = 0 \,.
\label{eq-1}
\end{equation}

Equation~\eq{eq-1} is well known in particle physics as it determines zero eigenmodes of a fermion field in a background of an abelian vortex~\cite{Jackiw:1981ee}. The magnetic flux of the abelian vortex is equal to the quantized vorticity number $n$. There are exactly $|n|$ isolated, linearly-independent, zero-energy bound states. These bound states are topologically protected by index theorems. For the sake of convenience here we repeat below the derivation of Ref.~\cite{Jackiw:1981ee}.

We represent $\psi = e^{-\sigma^3 f} \widetilde{\psi}$ and rewrite the Hamiltonian in the polar coordinates $r, \theta$ in the transverse plane of ${\bs x}_\perp = (x_1,x_2)$ using
\beqn
x_1 & = & r \, {\rm cos}\, \theta, \quad x_2 = r \, {\rm sin}\, \theta, \\
\hat{p}_r & = & - i \partial_r, \quad\ \ \hat{p}_\theta = - \frac{i}{r} \partial_\theta,
\eeqn
and the radial sigma matrices:
\begin{equation}
\sigma^r = \left(\begin{array}{cc} 0 & e^{-i\theta}\\e^{i\theta}& 0\end{array} \right), \quad \sigma^\theta = \left(\begin{array}{cc} 0 & -i e^{-i\theta}\\i e^{i\theta}& 0 \end{array}\right)
\end{equation}
Then equation for the function $\widetilde{\psi}$ is $\widetilde{\cal H}^{(R)}_\bot \widetilde{\psi} = 0$, where
\begin{eqnarray}
\widetilde{\cal H}^{(R)}_\bot &\approx &\phi(r,\theta) + \sigma^r  \hat{p}_r + \sigma^{\theta} \hat{p}_\theta\,,
\end{eqnarray}
or
\begin{equation}
\widetilde{\cal H}^{(R)}_\bot
\, \approx\,
\sigma^1 \left(\begin{array}{cc} e^{i\theta}{\cal H}^{(R)}_+ & \phi(r,\theta) \\ \phi(r,\theta) & e^{-i\theta}{\cal H}^{(R)}_-\end{array} \right), \quad {\cal H}^{(R)}_- = \Big[{\cal H}^{(R)}_+\Big]^{\dag}
\end{equation}
with
\begin{eqnarray}
{\cal H}^{(R)}_\pm &\approx & \hat{p}_r  \pm i  \hat{p}_\theta   \label{HS}
\end{eqnarray}

In the absence of the electric potential $\phi(r,\theta)$ the zero modes (if they exist) have a definite value of the spin projection $s = \pm 1/2$ on the $z$ axis. At large $r$ the corresponding coordinate parts of their wave functions satisfy the relations
\begin{equation}
(\hat{p}_r  \pm i  \hat{p}_\theta)\widetilde{\psi}_{\pm}^{(m)} = 0\,. \label{eq0}
\end{equation}
Next, we chose
\beqn
f(r,\theta) = \int_0^r \Hat{\Phi}(r,\theta)\frac{dr}{r}\,,
\eeqn
so that the angular $\theta$-component of the gauge potential~\eq{eq:A:if} gets the following form:
\beqn
A_\theta = \frac{\hat{\Phi}(r,\theta)}{r}\,.
\eeqn
The axial symmetry of the problem implies that at large distances $r$ the function $\hat{\Phi}(r,\theta)$ is independent of the polar angle $\theta$. Therefore, at large $r$ the function
$\hat{\Phi}(r)$ is the magnetic flux within the circle $S_r$ of radius $r$:
\begin{equation}
\hat{\Phi}(r) = \frac{1}{2\pi} \int_{S_r} \frac{1}{2} \epsilon_{ijk}dx^j\wedge  dx^k {H}^i(x,y)\,,
\end{equation}
where the surface $S_r$ belongs to the plane which is orthogonal to the dislocation. We come to the following solutions of Eq.~\eq{eq-1} for the
right-handed zero modes~\cite{Ansourian}:
\begin{equation}
\psi^{(m)}_\pm(r,\theta) \sim r^{m } e^{\pm i m \theta \mp \int_0^r \Hat{\Phi}(r,\theta)\frac{dr}{r}}\,,
\label{psim}
\end{equation}
where the integer $m$ is the angular quantum number.

The solutions~\eq{psim} are localized in a small vicinity of the dislocation core provided that the angular quantum number satisfies the following condition
\begin{equation}
m - 2 s \Hat{\Phi} < - 1,
\label{eq:condition:1}
\end{equation}
where
\begin{equation}
\hat{\Phi} \equiv \frac{\Phi}{\Phi_0} = \lim_{r \to \infty} \Hat{\Phi}(r,\theta)\,,
\label{eq:Phi:hat}
\end{equation}
is the total flux of the intrinsic magnetic field ${\bs H}$ normalized by the elementary magnetic flux~\eq{eq:Phi:0}. If Eq.~\eq{eq:condition:1} is satisfied then the corresponding probability distribution is convergent at large $r$:
\beqn
\int_{\xi}^{\infty} rdr d\theta |\psi|^2 = 1\,.
\label{eq:normalization:psi2}
\eeqn
Notice that the ultraviolet cutoff $\xi$ is of the order of the lattice constant~$a$.

In addition, there exist two solutions of Eq. (\ref{eq0}), which may not be normalized and which have their maxima at the dislocation core provided that
\begin{equation}
m = [2 s \hat{\Phi}(\infty)],
\end{equation}
where $[2s \Hat{\Phi}]$ is the integer part of $2s \Hat{\Phi}$, which is the maximal integer number that is not larger than $2s\Hat{\Phi}$.

The probability distributions of the considered solutions are convergent at small $r$ for $m\ge 0$. Therefore, in the absence of both the nontrivial vielbein and the electric potential, the zero modes  that are not singular at $r\rightarrow 0$ and are not localized on the boundaries of the system, should satisfy
$0 \le m  \le   2 s \Hat{\Phi}(\infty)$. Such modes exist for $s \Hat{\Phi}(\infty) > 0$ and are enumerated by the values of orbital momentum
\begin{equation}
m = 0, ..., [ 2s \Hat{\Phi}],
\end{equation}
We neglected in this derivation the potential $\phi$. However, it is localized at the dislocation. Therefore, the zero modes in the presence of electric potential (if they exist) have the form of Eq. (\ref{psim}) at $r \gg a$.
Recall, that Eq. (\ref{SHe_3sW2}) works for the momenta of quasiparticles much smaller than $1/a$. Therefore, the solutions of Eq. (\ref{eq-1}) localized at the dislocations, presumably, do not represent physical zero modes. The only solution that remains is the one with
\begin{equation}
m = [ 2s \Hat{\Phi}],\quad s = \frac{1}{2}{\rm sign}\, \hat{\Phi}
\end{equation}
Fortunately, the field $\phi$ cannot affect the energy of this solution because
the probability density corresponding to this solution of Eq. (\ref{eq0}) is dominated by the distances far from the dislocation core, so that we can neglect completely the region of the dislocation core. The vielbein  for this solution also gives small corrections  compared to the contribution of emergent magnetic field. Therefore, the strong gravity and the potential $\phi$ at $r \sim \xi$ cannot affect the main properties of this solution: it certainly survives as the zero mode and still has the definite value of the projection of spin to the $z$ axis.

In the case of edge or mixed dislocation we should take into account the appearance of a nonzero third component of the emergent gauge field:
\beqn
A_3 ({\bs x}_\perp)\approx \frac{1}{r^2}\Big(\beta_1 {\bs b}_\bot {\bs x}_\bot +  \beta_2 \epsilon_{3ij} b^i_\bot x^j_\bot\Big)\,.
\eeqn
Then far from the dislocation core one gets
\begin{equation}
\widetilde{\cal H}^{(R)}_\bot  = \sigma^1 \left(\begin{array}{cc} e^{i\theta}{\cal H}^{(R)}_+ & \nu A_3(r,\theta) \\ - \nu A_3(r,\theta) & e^{-i\theta}{\cal H}^{(R)}_-\end{array} \right)
\end{equation}

Perturbative corrections to the eigenenergy due to the presence of $A_3$ may be nonzero, in principle, but for the mode with $m = [2 s \Hat{\Phi}(\infty)]$ those corrections may be neglected because all integrals are dominated by the regions with $r\rightarrow \infty$ while $ A_3 \sim 1/r$ at large distances.

Thus we come to the conclusion, that the only zero mode existing around the dislocation is the one with
\beqn
m = [2 s \Hat{\Phi}(\infty)]\,, \qquad s = \frac{1}{2}{\rm sign}\,\Hat{\Phi}(\infty)\,.
\label{eq:ms:zero:R}
\eeqn

The zero mode~\eq{psim}, \eq{eq:ms:zero:R} of the transverse Hamiltonian ${\cal H}^{(R)}_\bot$ corresponds to the zero mode of the full Hamiltonian ${\cal H}^{(R)}$ provided the longitudinal momentum is zero $p_3 = 0$. At the same time it corresponds to a linear branch of spectrum of the full Hamiltonian ${\cal H}^{(R)}$
with the corresponding dispersion law:
 \begin{equation}
 {\cal E}^{(R)} \approx v_F \nu^{2/3} {\rm sign}(\Hat{\Phi}) \, p_3 \,.
 \label{ER}
 \end{equation}
This branch crosses zero energy level at $p_3 = 0$.

Similar considerations can also be applied to the left-handed Hamiltonian ${\cal H}^{(L)}$, where the only physical zero mode of the corresponding transverse part ${\cal H}^{(L)}_\bot$ is
\begin{equation}
\psi^{(m)}_{2s}(r,\theta) \sim r^{m } e^{ i 2 s m \theta - 2 s \int_0^r \Hat{\Phi}(r)\frac{dr}{r}}\,,\label{psiml}
\end{equation}
with the quantum numbers
\begin{equation}
m = [ 2s \Hat{\Phi}],\quad s = \frac{1}{2}{\rm sign}\, \hat{\Phi}\,.
\label{eq:ms:zero:L}
\end{equation}
This mode corresponds to the branch of spectrum with the dispersion
 \begin{equation}
{\cal E}^{(L)} \approx - v_F \nu^{2/3} {\rm sign}(\Hat{\Phi}) \, p_3\,.
\label{EL}
\end{equation}

The right-handed and left-handed fermionic modes propagate along the dislocation with the velocity
\beqn
v_R = - v_L = v_F \nu^{2/3} {\rm sign}(\Hat{\Phi}) \,,
\label{eq:vR:vL}
\eeqn
which is nothing but the corresponding component of the anisotropic Fermi velocity. Thus, the right-handed massless quasiparticle propagates up or down along the dislocation depending on the sign of the flux $\Phi$. The left-handed mode always propagates in the opposite direction compared to the right-handed mode.

Sure, these mode may be coupled with the counter-propagating zero modes, for example, by the intra - valley and/or inter - valley impurity scattering. This effect is neglected here. But this is what is also neglected always in the Dirac semimetals, when one speaks of the massless Dirac fermions emergent in those materials. From the theoretical side the gapless nature of the emergent Dirac fermions in Dirac semimetals should be  protected by momentum space topology: the symmetry - protected topological invariant in momentum space should exist that protects the Dirac cones (see \cite{Volovik2003}). The corresponding symmetry is specific for the crystals of the semimetals. For example, for the case of Na$_3$Bi the effective model (that follows from the microscopic electronic model) is given in \cite{ref:kzDirac}). This model utilizes the time reversal,
inversion, and $D^4_{6h}$ symmetries. Their composition (possibly, supplemented by something) should, in principle, protect the Dirac cone for the case of Na$_3$Bi. But the extensive theoretical study of this issue is out of the scope of the present paper.

Anyway, the symmetry responsible for the gapless nature of the modes localized around the dislocation is the same, which is responsible for the gapless nature of the bulk quasiparticles. This symmetry is not exact, and it is indeed broken softly  by the impurity scattering and also by the collisions with the change of chirality. The corresponding parameter is the mean free path, which serves here as the infrared cutoff of the theory and as the typical size of the region, where the zero modes of the present section are localized (see discussion in the forthcoming sections).

Notice that Eqs.~\eq{ER} and \eq{EL} were derived in the assumption that the magnetic fluxes of the emergent magnetic field for the right-handed $\Phi_R$ and the left-handed $\Phi_L$ quasiparticles are the same, $\Phi_R = \Phi_L \equiv \Phi$. However, if in a Dirac semimetal the constants $\beta_{ijk}$ differ for the left-handed and the right-handed fermions, then the corresponding gauge fields~\eq{DiracPosition1} are also different, and in this case the magnetic flux entering Eq. (\ref{ER}) will be different from the flux in Eq. (\ref{EL}).

\section{Chiral anomaly in Dirac semimetals along the dislocation}

\label{SectCAWDS}

In the presence of external electric field ${\bs E}$ the states that correspond to the described above zero modes flow in the correspondence with the following equation:
\begin{equation}
\langle \dot{p}_3 \rangle = E_3 \,.
\end{equation}

Now let us take into account, that the studied model has the infrared cutoff $1/\xi_0$, where $\xi_0 \gg a$. Then the zero modes of ${\cal H}^{(R)}_\bot$ and ${\cal H}^{(L)}_\bot$, and the corresponding branches of spectrum of the propagating modes of the full Hamiltonians ${\cal H}^{(R)}$ and ${\cal H}^{(L)}$
obey the following properties:

\begin{enumerate}

\item{} The propagating fermion modes are not localized at the dislocation core. Instead, the region of space around the dislocation of size $\xi_0$ dominates, where $1/\xi_0$ is the infrared cutoff of the theory.

\item{} The propagating fermion modes have the definite value of the spin projection on the dislocation axis: $s = \frac{1}{2}{\rm sign}\,\Hat{\Phi}$. The corresponding branch of spectrum for the right- and the left-handed fermions is given, respectively, by the following dispersion relations:
\beqn
{\cal E}_{R/L}(p_3) = \pm 2s v_F \nu^{2/3} p_3\,.
\label{eq:E:RL}
\eeqn

\item{} The propagating mode appears for any dislocations including those ones, in which the magnetic flux $\Hat{\Phi}$ is smaller than unity.

\end{enumerate}

The total production of the right-handed quasiparticles per unit length of the dislocation is given by:
\begin{equation}
\dot{q}_R =  \frac{{\bs E}{\bs n}}{2\pi}   {\rm sign}\, \Hat{\Phi},\label{Aq}
\end{equation}
where the unit vector $\bs n$ is directed along the dislocation. In the following we assume for simplicity, that the signs of the emergent fluxes $\hat{\Phi}$ experienced by the right-handed and the left-handed fermions coincide in the Dirac semimetal. Therefore, the production of the left-handed quasiparticles in Dirac semimetal is given by
\beqn
\dot{q}_L=-\dot{q}_R\,.
\label{Aq2}
\eeqn

The production  of the quasiparticles may be written as the anomaly in their currents
\beqn
{j}^{\mu}_L & = & |{\bs e}(x)| J^{\mu}_L(x)\,, \qquad {j}^{\mu}_R = |{\bs e}(x)| J^{\mu}_R(x)\,,\\
{\bs j} & = & {\bs j}_R + {\bs j}_L\,, \qquad\quad\ \, {\bs j}_5 = {\bs j}_R - {\bs j}_L\,,
\eeqn
where the covariant currents are defined according to Eqs. (\ref{JR}) and (\ref{JL}). In a local form the anomaly may be expressed as follows:
\beqn
\langle \partial_\mu {\bs j}^{\mu}_5(x) \rangle &=& \frac{{\bs E}{\bs n}}{\pi} \, f_0(x_\perp) \, {\rm sign}\, \Hat{\Phi}\,,  \label{CA}
\eeqn
where the function $f_0$ can be read from Eqs.~\eq{psim}, \eq{psiml}:
\beqn
f_0(x_\perp) & = & \frac{\exp\left( - |x_\perp|/\xi_0 \right)
\Big(\frac{x_\perp}{\xi_0}\Big)^{-2(|\Hat{\Phi}|-[|\Hat{\Phi}|])}}{2\pi \xi_0^2 \Gamma(-2|\Hat{\Phi}|+2[|\Hat{\Phi}|]+2)}\,. \quad
\label{f0}
\eeqn
This function is normalized in such a way, that
\beqn
2\pi \int^{\infty}_{0} r dr f_0(r) = 1\,.
\eeqn
The quantity $1/\xi_0$ has the meaning of the infrared cutoff of the theory
and the factor $\exp\left( - |x_\perp|/\xi_0 \right)$ appears as the infrared regulator. We imply that the size of the semimetal sample is much larger than the infrared cutoff $\xi_0$, while $\xi_0$ is much larger than the size of the dislocation core $\xi \sim a$, $\xi_0\gg \xi$. Thus the chiral anomaly due to the zero mode with $m = [|\Hat{\Phi}|]$ is localized within the tube of size $\xi_0$ centered at the dislocation.

In our derivation we follow the procedure first applied by Nielsen and Ninomiya in \cite{Nielsen:1983rb}. This derivation is now standard and it was repeated many times. In particular, the derivation of Eq. (\ref{Aq}), (\ref{Aq2}) may be found in Sect. 18.1.1 - 18.1.3. of \cite{Volovik2003}. Eq. (\ref{CA}) is the local form of Eqs. (\ref{Aq}), (\ref{Aq2}). It appears as the local violation of the continuity equation that follows from the appearance from vacuum of the state with the wave function of Eq. (\ref{psiml}).

It is worth mentioning, that for a single dislocation, if for some reasons the contributions to the emergent magnetic flux of the dislocation due to $\beta_{ijk}$ in Eq.~(\ref{DiracPosition1}) may be neglected, the typical values of the Burgers vector are such that $|\Hat{\Phi}|< 1$. Therefore, according to Eqs.~\eq{eq:ms:zero:R} and \eq{eq:ms:zero:L}, for a single dislocation the zero mode corresponds to the angular momentum $m = 0$.

We may rewrite the expression for a chiral anomaly caused by a single dislocation in a Dirac semimetal as follows
\begin{eqnarray}
\langle \partial_\mu j^{\mu}_5(x) \rangle &=& \frac{{\bs E}{\bs B}_{}}{2\pi^2}\,,
\label{CADS}
\end{eqnarray}
where the effective magnetic field ${\bs B}$ responsible for the chiral anomaly is given by
\begin{eqnarray}
{\bs B}_{}({\bs x}_\perp) &=&  2\pi {\bs n} \,   f_0({\bs x}_\perp) \, {\rm sign}\, \Hat{\Phi}\,,
\label{BCADS}
\end{eqnarray}
where the function $f_0$ is given by Eq.~\eq{f0}.

It is worth mentioning, that the above consideration refers only to those branches of spectrum, which are described by the low energy effective field theory. In the presence of electric field the pumping of the quasiparticles from vacuum may also occur at another branches of spectrum. Ideally, this pumping process should be treated with the help of a microscopic theory and is out of the scope of the present paper.

\section{The case of dislocation directed arbitrarily}
\label{SectAnis}

In this section we consider the dislocation directed arbitrarily. Without loss of generality we consider the dislocation directed along an axis, which belongs to the $(yz)$ plane. The angle between the dislocation and the $z$ axis is denoted by $\varphi$. Let us rotate the reference frame in such a way, that the $z$ axis is directed along the dislocation. In the new reference frame the tensor $\hat{f}$ has the form:
\begin{equation}
\hat{f} = \left(\begin{array}{ccc} \nu^{-1/3} & 0 & 0 \\
0 & \nu^{-1/3} \, {\rm cos}\, \varphi & \nu^{-1/3} \, {\rm sin}\, \varphi\\
0 & -\nu^{2/3} \, {\rm sin}\, \varphi & \nu^{2/3} \, {\rm cos}\, \varphi
\end{array} \right)
\end{equation}

Let us apply the transformation of spinors $\psi \rightarrow e^{i \frac{\alpha}{2}\sigma^3}$ with ${\rm tg}\,\alpha = \nu \, {\rm tg}\, \varphi$. In the transformed frame the tensor $\hat{f}$ is modified:
\begin{widetext}
\begin{equation}
\hat{f} = \left(\begin{array}{ccc} \nu^{-1/3} & 0 & 0 \\
0 & \nu^{1/3}\sqrt{\nu^{-4/3} \, {\rm cos}^2\, \varphi + \nu^{2/3} \, {\rm sin}^2\, \varphi} &
\frac{(1-\nu^2)\,{\rm sin}\,2\varphi}{2\nu\sqrt{\nu^{-4/3} \, {\rm cos}^2\, \varphi + \nu^{2/3} \, {\rm sin}^2\, \varphi}}\\
0 & 0 & \frac{1}{\sqrt{\nu^{-4/3} \, {\rm cos}^2\, \varphi + \nu^{2/3} \, {\rm sin}^2\, \varphi}}
\end{array} \right)
\end{equation}
\end{widetext}

The one-particle Hamiltonian for the right-handed fermions becomes as follows
\begin{eqnarray}
{\cal H}^{(R)}  &=& v_F\hat{f}^3_3 \sigma^3  \hat{p}_3  +  v_F\hat{f}^2_3 \sigma^2  \hat{p}_3 + v_F \hat{f}^1_1 {\cal H}^{(R)}_\bot
\end{eqnarray}
with
\begin{eqnarray}
{\cal H}^{(R)}_\bot & \approx &  \sigma^1  (\hat{p}_1 -  A_1) + \frac{\hat{f}_2^2}{\hat{f}^1_1}\sigma^2 (\hat{p}_2 -  A_2) \nonumber\\ && - \frac{\hat{f}_3^3}{\hat{f}^1_1}\sigma_3 A_3(x_\bot)+\phi(x_\bot)
\end{eqnarray}

Now we perform the coordinate transformation
\beqn
y \rightarrow \frac{\hat{f}_2^2}{\hat{f}^1_1} y\,, \qquad A_y \rightarrow \frac{\hat{f}_1^1}{\hat{f}^2_2}  A_y\,,
\eeqn
and notice that the equation for the zero mode of the Hamiltonian ${\cal H}_\bot$ becomes the same as the one discussed in Section~\ref{SectZMD}. Thus we arrive at the expression for the anomaly in quasiparticle current of Eq. (\ref{Aq}). The resulting expression for the chiral anomaly in the Dirac semimetal is again given by Eq. (\ref{CADS}).

\section{Chiral density and chiral chemical potential around the dislocation in the presence of electric field}

\label{SectCDS}

The Dirac semimetal possesses two cones, each of which hosts one right-handed and one left-handed Weyl fermion. Since the processes operating in these two cones are equivalent, we concentrate on one  cone hereafter taking into account the fact of the double degeneracy later.

The evolution of the local chiral density around the dislocation is governed by (i) the generation of the local chiral charge due to quantum anomaly at the dislocation given by Eqs.~\eq{CADS} and \eq{BCADS}, (ii) the spatial diffusion of the chiral charge and (iii) the dissipation of the chiral charge density~\eq{eq:rho5}.
In  {the other words, in the presence of the external electric field, the zero modes, distributed around the dislocation and propagating along the dislocation, accumulate the chiral charge}. The accumulated chiral charge diffuses (due to, basically, thermal diffusion and scattering) and also dissipates (due to  {the chirality-changing processes}) around the dislocation. We estimate these effects below.

At the distances larger than the size of the dislocation core $r \gg \xi \sim a$ we may neglect the presence of the  {emergent magnetic field ${\bs H}$ of the dislocation. Therefore, in order to relate the chiral chemical potential
$\mu_5$
with the chiral density
$\rho_5$
at finite temperature $T$ we use an approximation, in which the relevant modes of the quasiparticles are the plane waves of the continuous spectrum. Thus, }
we neglect gauge field completely and calculate the thermodynamical potential (see \cite{ZrTe5} and references therein):
\begin{equation}
\Omega = T \sum_{s=\pm 1}\sum_{c = \pm 1} \int \frac{d^3 p}{(2\pi)^3}\,{\rm log}\, \Bigl(1+e^{-\frac{\omega_{p,s}+ c \mu_5}{T}}\Bigr)\,,
\label{eq:Omega:1}
\end{equation}
where {the quantity $c = \pm 1$ labels right- and left-handed chiralities,}
$s \,{= \pm 1}$ is the projection of spin
{(multiplied by two)}
to
{the auxiliary}
vector $k_a ({\bs p}) = \hat{f}^i_a p_i$,
while $p_i$ is momentum of the quasiparticle. The chiral chemical potential $\mu_5$ is the difference between the chemical potentials associated with the fermions of right-handed and left-handed chiralities:
\beqn
\mu_5 = \frac{1}{2} \left( \mu_R - \mu_L \right)\,.
\eeqn

In Eq.~\eq{eq:Omega:1} the dispersion of the quasiparticles in terms of the vectors $\bs p$ and $\bs k$ is as follows:
\begin{equation}
\omega_{p,s} = c\, s \, v_F\,\sqrt{\hat{f}^i_a\hat{f}^j_ap_ip_j} = c\, s \,v_F\,|{\bs k} \, {({\bs p})}|\,,
\label{eq:omega:ps}
\end{equation}
where $v_F$ is the Fermi velocity~\eq{eq:vR:vL} which enters the dispersion relation for the chiral fermions~\eq{eq:E:RL}.
The matrix $\hat f$ is given in Eq.~\eq{eq:hat:f}.
For the momentum parallel to the $z$ axis, ${\bs p} = (0,0,p_3)$, Eq.~\eq{eq:omega:ps} leads to the dispersion~\eq{eq:E:RL}.
In our calculation we work (following, e.g., Ref.~\cite{ZrTe5}) in the adiabatic approximation assuming that the chiral chemical potential is a slowly varying function of space and time.

Since the determinant of the matrix $\hat{f}$  {is equal} to unity, we get for the  {thermodynamical} potential~\eq{eq:Omega:1}:
\beqn
\Omega = T \sum_{s=\pm 1}\sum_{c = \pm 1} \int \frac{d^3 k}{(2\pi)^3}\,{\rm log}\, \Bigl(1+e^{-\frac{c\, s \,v_F\,|{\bf k}|+ c \mu_5}{T}}\Bigr). \qquad
\label{eq:Omega:def}
\eeqn
The chiral density is given by the derivative of the thermodynamical potential
$\Omega$ with respect to chiral chemical potential $\mu_5$:
\begin{equation}
\rho_5  \,{= \frac{\partial \Omega}{\partial \mu_5}}
= -\frac{1}{2\pi^2} \sum_{s=\pm 1}\sum_{c = \pm 1} \int \frac{c\, {k^2d k}}{1+e^{\frac{c\, s \,v_F\,k+ c \mu_5}{T}}}
\end{equation}
Obviously, for $\mu_5=0$ the chiral density vanishes. For $|\mu_5| \ll T$ we may evaluate the term in $\rho_5$ linear in $\mu_5$ differentiating the last expression with respect to $\mu_5$:
\begin{eqnarray}
\rho_5 &\approx & \frac{\mu_5}{2\pi^2\,T} \sum_{s=\pm 1, c = \pm 1}  \int   \frac{e^{\frac{c\, s\, v_F\,k}{T}}k^2 dk }{\Big(1+e^{\frac{c\, s\,v_F\,k}{T}}\Big)^2}\nonumber\\
&=& \frac{2\mu_5}{\pi^2\,T}  \int   \frac{k^2 dk }{\cosh^2\bigl({\frac{v_F\,k}{2T}\bigr)}} = \frac{\mu_5\,T^2}{3\,v_F^3 } \,.
\label{eq:rho5}
\end{eqnarray}

Equation~\eq{eq:rho5} is valid provided  {that certain conditions are satisfied. First, the size of the dislocation core $\xi \sim a$ (where the low-energy physics, and, consequently, the thermodynamic relation~\eq{eq:rho5} both become inapplicable), should be small compared to the wavelength of the typical thermal momentum $\lambda_T \sim 1/p_T \sim v_F/T$ that contributes to relation~\eq{eq:rho5}. According to our estimates (see below, Section~\ref{SectNUM}) this condition is satisfied even for the room temperature $T \sim 300 \,\mbox{K}$ (with corresponding $\lambda_T \sim 4\times 10^{-8}\,\mbox{m}$) because $\xi \sim 10^{-9}\,\mbox{m}  \ll \lambda_T$ according to Eq.~\eq{eq:estimate:xi}. Second,} the magnetic field of the dislocation should not affect considerably the plane waves contributing to Eq.~\eq{eq:rho5}. To this end one can consider a wavefunction of a particle that circumferences the dislocation, and compare the contributions to its phase coming from the magnetic field and from the usual kinetic factor $e^{i p x}$.  The former reaches its maximum at $r = \xi_0$, being equal to the total flux $\delta \phi_\Phi = \Phi \sim 2 \pi$ while the later can be estimated as $\delta \phi_T = 2 \pi p_T \xi_0$. Thus, the second condition requires $\delta \phi_\Phi \ll \delta \phi_T$ or $\lambda_T \ll \xi_0$ which is also satisfied according to Eq.~\eq{eq:estimate:xi0}.

The nonconservation of the axial charge can conveniently be written in the following form (see \cite{ZrTe5}):
\beqn
\frac{d \rho_5}{d t} + {\bs \nabla} {\bs j}_5 = - \frac{\rho_5}{\tau_V}   + \frac{1}{2 \pi^2} {\bs B} {\bs E}\,,
\label{eq:drho5}
\eeqn
where the first term in the right hand side corresponds to the dissipation of the chiral charge with the rate given by the chirality-changing scattering time $\tau_V$ while  the second term describes the generation of the chiral charge due to the quantum anomaly around the dislocation.
In fact, this equation appears also in the chiral kinetic theory developed relatively (see, for example, \cite{Stephanov:2012ki,Chen:2015gta}) when the kinetic equations are taken in the relaxation time approximation (also known as $\tau$ approximation).

The chiral current,
\beqn
{\bs j}_5 = - D_5 {\bs \nabla} \rho_5 \,,
\label{eq:j5:D5}
\eeqn
is given by the diffusion of the chiral charge $\rho_5$ with the corresponding diffusion constant $D_5$. This equation implies that the chiral charge is able to diffuse. We do not have the microscopic description of the diffusive nature of the chiral charge, and at the present level it is considered here as the hypothesis.

 We assume that the Dirac semimetal has zero usual chemical potential for the Dirac quasiparticles, $\mu = 0$. Moreover, we consider a linear approximation so that  the transport effects, which are discussed here, do not generate a nonzero $\mu$.
Substituting Eq.~\eq{eq:j5:D5} into Eq.~\eq{eq:drho5} one gets the following equation for the chiral charge density:
\beqn
\frac{d \rho_5}{d t} = - \frac{\rho_5}{\tau_V}  + D_5 \Delta \rho_5 + \frac{1}{2 \pi^2} {\bs B} {\bs E}\,.
\label{eq:drho5:2}
\eeqn

In the constant electric field, $d {\bs E} / d t = 0$, the chiral charge $\rho_5$ relaxes towards equilibrium $d \rho_5/d t = 0$ at late times $t \gg \tau_V$. The equilibrium chiral charge density is given by a solution of Eq.~\eq{eq:drho5:2} with vanishing left hand side:
\beqn
\rho_5(x) = \frac{1}{2 \pi^2 D_5} \int d^3 y \, G^{(3)}(x - y;\lambda) \bigl({\bs B}(y) \cdot {\bs E}(y)\bigr), \qquad
\label{eq:rho5:EB}
\eeqn
where
\beqn
\left( - \Delta + L^{-2}_V \right) G^{(3)}({\bs x} - {\bs y};\lambda) = \delta({\bs x} - {\bs y})\,,
\label{eq:G:3d}
\eeqn
is the three-dimensional Green's function and
\beqn
L_V = \sqrt{D_5 \tau_V}\,,
\label{eq:L:V}
\eeqn
is a characteristic length which controls spatial diffusion of the chiral charge.

Working in {linear approximation we consider weak electric field ${\bs E}$}, so that the chiral imbalance can always be treated as a small quantity, $\mu_5 \ll T$, so that the linear approximation in Eq.~\eq{eq:rho5} is justified.
In the absence of the usual chemical potential $\mu$, one gets from Eq.~\eq{eq:rho5} the following relation between the chemical potential and the chiral charge density~\eq{eq:rho5:EB}:
\beqn
\mu_5 (x) = \frac{3 \,v_F^3}{T^2} \rho_5 (x)\,.
\label{eq:mu5:rho5}
\eeqn
Thus we see, that the dislocation produces the chiral charge which spreads around the dislocation, effectively creating an excess of the chiral chemical potential at the characteristic distance $L_V$ from the dislocation axis. Notice, that this chiral chemical potential corresponds to a single Dirac point with a pair of Weyl fermions.

In the above derivation we neglect the gradient of temperature. This may be done for sufficiently small external electric field, when the temperature remains almost constant at the characteristic length of the problem that is $\xi_0$.  Very roughly, in equilibrium the heat generated by electric field $\sim \sigma E^2$ (where $\sigma$ is the total conductivity that includes Ohmic contribution) should be equal to the divergence of the  heat flow $\kappa \nabla T$ (where $\kappa$ is the thermal conductivity). For our estimate we use the Wiedemann-–Franz law $\kappa \sim \sigma T$. This gives for the characteristic length $\xi_T$ (at which the temperature is changed considerably):
\begin{equation}
\frac{1}{\xi_T^2}\sim \frac{\Delta T}{T} \sim \frac{E^2}{T^2}
\end{equation}
According to our estimates (see below Sect. \ref{SectNUM}) at room temperatures the condition $\xi_T \gg \xi_0$ leads to $|E| \ll 1 $ V$/$cm. This condition provides, that temperature remains constant within the region of size $\xi_0$ around the dislocation. However, temperature may vary within the whole semimetal sample if its size is much larger than $\xi_0$.

Now let is consider practically interesting case when the dislocation is a strait line centred at the origin, $x_1= x_2 =0$ and directed along the $x_3$ axis. The dislocation induces the intrinsic magnetic field
\beqn
{\bs B}(x_\perp) = B_z(x_\perp) {\bs n}\,,
\eeqn
which is also directed along the $x_3 \equiv z$ axis (here ${\bs n}$ is the unit vector in $z$ direction). The intrinsic magnetic field is a function of the transverse coordinates $x_\perp = (x_1,x_2)$ which takes nonzero values in a (small) core of the dislocation. In our model approach we consider the field given by Eqs.~\eq{f0} and \eq{BCADS}
\beqn
B_z(x_\perp) = {\rm sign}\, \Phi \, \frac{\exp\left( - |x_\perp|/\xi_0 \right)\Big(\frac{x_\perp}{\xi_0}\Big)^{-2(|\Hat{\Phi}|-[|\Hat{\Phi}|])}}{ \xi_0^2 \Gamma(-2|\Hat{\Phi}|+2[|\Hat{\Phi}|]+2)},\quad\
\label{eq:B:Gauss}
\eeqn
where
\beqn
\alpha = |\hat{\Phi}|-[|\hat{\Phi}|]\,,
\eeqn
is the fractional charge of the normalized flux $|\hat \Phi|$. The effective magnetic field~\eq{eq:B:Gauss} is distributed around the dislocation with the characteristic length $\xi_0$ that is much larger than the size $\xi$ of the dislocation core (the latter is of the order of a few lattice spacings $a$). In Fig.~\eq{fig:B} we show the field~\eq{eq:B:Gauss} for a few values of the fractional part  of the flux $\alpha$.
\begin{figure}[!thb]
\begin{center}
\vskip 3mm
\includegraphics[scale=0.57,clip=true]{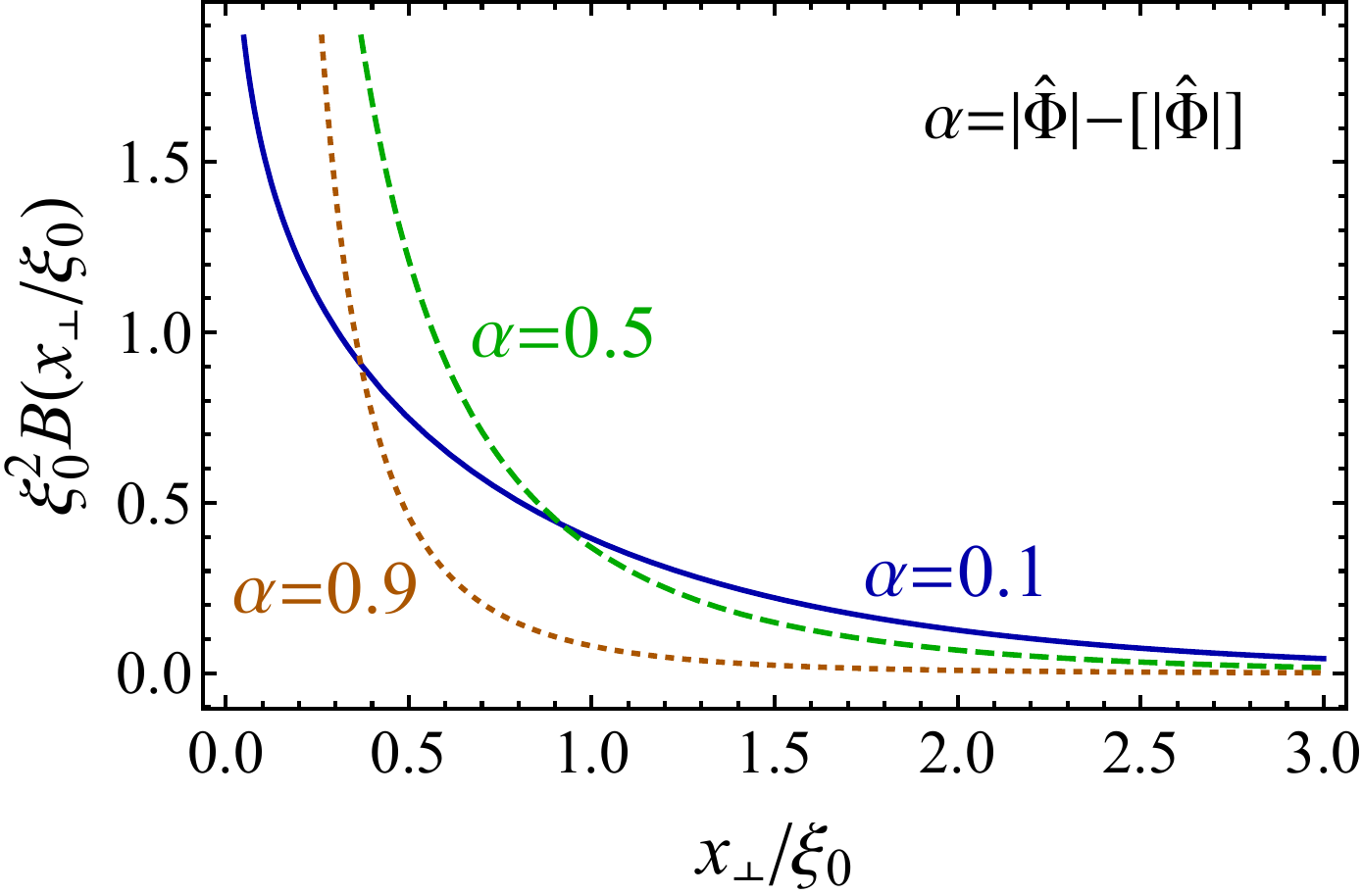}
\end{center}
\vskip -3mm
\caption{The effective intrinsic magnetic field $B_z(x_\perp)$ in Eq.~\eq{eq:B:Gauss} vs. the distance from the dislocation core $x_\perp$ plotted for a few values of
the fractional part of the absolute value of the normalized flux~\eq{eq:Phi:hat} $\alpha =0.1,\,0.5,\,0.9$.}
\label{fig:B}
\end{figure}

This {\it effective} magnetic field carries the unit of the elementary flux~\eq{eq:Phi:0}: $\int d^2 x_\perp B(x_\perp)= 2\pi\, {\rm sign} \, \Phi$. As we have discussed earlier, this effective magnetic field is associated with the propagating zero modes bounded at the dislocation. In Eq.~\eq{eq:B:Gauss} the total flux $\Phi$  of the intrinsic magnetic field ${\bs H}$ is of a geometrical origin. The flux is a quantity of the order of unity~\eq{eq:Phi:hat}, in terms of the elementary magnetic flux~\eq{eq:Phi:0}. For the straight dislocation~\eq{eq:B:Gauss} the axial anomaly generates the axial charge which spreads in the semimetal in the transverse directions according to the equilibrium formula~\eq{eq:rho5:EB}.

For the completeness we represent the numerical estimates for the encountered above constants that characterize the semimetal in Appendix B.

\section{Conclusions}

\label{SectConcl}

In this work we discussed certain effects of anomalies in the Dirac  semimetals Na$_3$Bi  and Cd$_3$As$_2$ caused by the dislocations in their atomic lattices. This chiral anomaly is operational without any external magnetic field unlike the conventional chiral anomaly that was discussed for Dirac semimetals, for example, in \cite{ZrTe5}. The dislocation appears as a source of the emergent magnetic field as it carries the emergent magnetic flux. The emergent flux gives rise to the zero mode of the one-particle Hamiltonian for the fermionic quasiparticles. The fermionic mode is a gapless excitation which propagates along the dislocation being localized in the area of the size $\xi_0$ around the dislocation (here $1/\xi_0$ is the infrared cutoff of the field theoretical approximation used in our approach). The length $\xi_0$ may also be identified with the mean free path of the quasiparticles. For example, for Cd$_3$As$_2$ it is of the order of $200\, \mu$m.  This propagating zero mode corresponds to the branch of spectrum of the quasiparticles with the spin directed along the magnetic flux of the dislocation.

In the presence of an external electric field the spectral flow along the zero-mode branch of the spectrum leads to the pumping of the quasiparticles from the vacuum. Since the right-handed and left-handed quasiparticles are produced with opposite rates, the pumping process corresponds to the chiral anomaly~(\ref{CADS}). The production rate of the chiral density is controlled by a scalar product of the usual (external) electric field ${\bs E}$ and the effective (internal) magnetic field ${\bs B}$ carried by the dislocation. One should stress the following subtle fact: the dislocation carries also the emergent magnetic field ${\bs H}$ [given in Eq.~\eq{PhiNa3Bi} for the example case of Na${}_3$Bi] which gives rise to the existence of the mentioned propagating zero mode. However, the emergent field ${\bs H}$ does not directly contribute to the chiral anomaly~(\ref{CADS}): it is the effective magnetic field ${\bs B}$ -- that is expressed via the density of the zero mode~\eq{f0} -- which enters the anomaly relation~(\ref{CADS}).

In {the} other words, the emergent magnetic field ${\bs H}$ with the flux~\eq{PhiNa3Bi} leads to appearance of the right-handed~\eq{psim} and the left-handed modes~\eq{psiml} localized at the dislocation and propagating along its axis. The external electric field ${\bs E}$, parallel to the dislocation axis, produces the chiral charge by pumping these modes form the vacuum at unequal (in fact, opposite) rates proportional to the scalar product ${\bs E} {\bs B}$. The process can be formulated via the chiral anomaly equation~\eq{CADS}, in which, however, the magnetic field ${\bs B}$ is expressed through the density of the wave functions of the mentioned zero modes (with ${\bs B} \neq {\bs H}$).

The chiral anomaly gives rise to a nonzero charge density localized around the dislocation axis with the characteristic localization length $L_V \sim
2\, \mu$m for Cd$_3$As$_2$. The slowly varying chiral density can be expressed via a (space-dependent) chemical potential.

In principle, there are various ways to create a Dirac semimetal with dislocations. In general, the growth of an atomic crystal may be organized in such a way, that the dislocations appear along the chosen direction with the chosen values of the Burgers vector. In addition, dislocations may also appear as a result of plastic deformations of the crystals~\cite{ref:dislocations}. This opens a possibility to observe the effects of the dislocation-induced anomaly experimentally. The chiral density that is formed around the dislocation in the presence of external electric field should affect transport properties of the semimetal. In order to calculate the corresponding observable quantities, this is necessary to use kinetic theory modified accordingly in order to take into account the appearance of the chiral density around the dislocations driven by chiral anomaly. However, the solution of this problem is out of the scope of the present paper.

\acknowledgments
The  part of the work of  M.A.Z. performed in Russia  was supported by Russian Science Foundation Grant No 16-12-10059 (Sections \ref{SectDWS}, \ref{SectEMD}, \ref{SectZMD}) while the part of the work made in France (Sections \ref{SectCAWDS}, \ref{SectAnis}, \ref{SectCDS}) was supported by Le Studium Institute of Advanced Studies.

\section*{Appendix A. Emergent gravity around the dislocation}
\label{SectEGD}

\vskip 3mm

 In this section we briefly consider emergent gravity around the dislocation. Let us represent the action for the right-handed fermion in the following way:
\begin{eqnarray}
S_R &=& \int\, d^3x\,dt \,  \bar{\Psi}(x,t)\Big[ |{\bs e}(x)|  e_0^0(x) i \partial_t  \nonumber\\ && -  |{\bs e}(x)|  e_a^k(x){\sigma}^a \circ (\hat P_k -  A_k)\Big] \Psi(x,t)\nonumber\\
&=& \int\, d^3x\,dt \,\bar{\widetilde{\Psi}}(x,t)\Big[  i \partial_t - {\cal H}^{(R)} \Big] \widetilde{\Psi}(x,t),
\end{eqnarray}
where $\hat P_k = - i \nabla_k$, ${\widetilde{\Psi}} = \sqrt{|{\bs e}(x)| e_0^0(x)}\,\Psi$, $a = 0,1,2,3$ and $k = 1,2,3$. The one-particle Hamiltonian is given by
\begin{eqnarray}
{\cal H}^{(R)}  &=&   {f}_a^k(x){\sigma}^a \circ (\hat P_k -  A_k),
\label{eq:HR}
\end{eqnarray}
where
\beqn
{f}_a^k(x)   =  \frac{ e_a^k(x)}{ e_0^0(x)}\,.
\label{eq:f:ak}
\eeqn
We used here the following chain of relations:
\begin{widetext}
\begin{eqnarray}
&&\int\, d^3x\,dt \,\left\{\frac{\bar{\widetilde{\Psi}}(x,t)}{\sqrt{|{\bs e}(x)|  e_0^0(x)}}\, |{\bs e}(x)|  e_a^k(x){\sigma}^a \partial_i \frac{\widetilde{\Psi}(x,t)}{\sqrt{|{\bs e}(x)|  e_0^0(x)}} - \left[\partial_i\frac{\bar{\widetilde{\Psi}}(x,t)}{\sqrt{|{\bs e}(x)|  e_0^0(x)}}\right]\, |{\bs e}(x)|  e_a^k(x){\sigma}^a  \frac{\widetilde{\Psi}(x,t)}{\sqrt{|{\bs e}(x)|  e_0^0(x)}}
\right\}
\nonumber\\
&=&\int\, d^3x\,dt \,\left\{\bar{\widetilde{\Psi}}(x,t)\frac{1}{{|{\bs e}(x)|  e_0^0(x)}}\, |{\bs e}(x)|  e_a^k(x){\sigma}^a \partial_i \widetilde{\Psi}(x,t) -
\Big[\partial_i\bar{\widetilde{\Psi}}(x,t)\Big]\, |{\bs e}(x)|  e_a^k(x){\sigma}^a  \frac{1}{{|{\bs e}(x)|  e_0^0(x)}} \widetilde{\Psi}(x,t)\right\}\\
& = &
\int\, d^3x\,dt \,\left\{\bar{\widetilde{\Psi}}(x,t)f_a^k(x)\, {\sigma}^a \partial_i \widetilde{\Psi}(x,t)
-\Big[\partial_i\bar{\widetilde{\Psi}}(x,t)\Big]\, {\sigma}^a  f_a^k(x) \widetilde{\Psi}(x,t)\right\}
\equiv 2 \int\, d^3x\,dt \,\bar{\widetilde{\Psi}}(x,t) \left[f_a^k(x)\, {\sigma}^a \circ \partial_i \right] \widetilde{\Psi}(x,t)
.\nonumber
\end{eqnarray}
\end{widetext}
 We represent ${f}_a^k(x)$ as follows
\begin{eqnarray}
{f}_a^k(x) & \approx & v_F\Big[\hat{f}_a^{k} - \hat{f}_b^k\delta  e^{b}_{a}(x)\Big]\nonumber\\
{f}_0^k(x) & \approx & - v_F\hat{f}_b^k\delta e^{b}_{0}(x),
\quad a,b,k = 1,2,3\,,
\label{eq:f:delta:e}
\end{eqnarray}
where the expressions for the small variations of the vierbein field $\delta e^\mu_a$ can be read off from Eq.~\eq{eg}.

The one-particle Hamiltonian for the right-handed fermions in the presence of a dislocation along the $z$ axis is given by
\begin{eqnarray}
{\cal H}^{(R)}  &=& v_F\nu^{2/3} \sigma^3  \hat{p}_3  - v_F \nu^{2/3} \sum_{a=0}^3
\sigma^a \delta  e^{3}_{a}\hat{p}_3
\nonumber\\
&& + v_F \nu^{-1/3}{\cal H}^{(R)}_\bot\,,
\label{eq:H:R}
\end{eqnarray}
where the transverse part of the Hamiltonian is
\begin{eqnarray}
{\cal H}^{(R)}_\bot & \approx & \sum_{a=1,2}
\Big[\sigma^a  \Bigl(\hat{p}_a - A_a({\bs x}_\perp) \Bigr)- \sum_{k=1,2}\sigma^a \delta  e^{k}_{a}({\bs x}_\perp)\circ \hat{p}_k \Bigr] \nonumber\\
&&  + \phi(x,y) - \nu \sigma^3  A_3({\bs x}_\perp)
\label{eq:H:R:transverse}\\
&&
- \sum_{k=1,2}\Bigl[\sigma^3 \delta  e^{k}_{3}({\bs x}_\perp) + \delta  e^{k}_{0}({\bs x}_\perp)\Bigr]\circ \hat{p}_k\,.\nonumber
\end{eqnarray}

In a general form, the dislocation-induced deformations of the vielbein field $\delta e^\mu_a$ in the Hamiltonian~\eq{eq:H:R}, \eq{eq:H:R:transverse} can be expressed via components tensor $\gamma^i_{jkl}$ of Eq.~\eq{eg} and the relations given in Eqs.~\eq{eq:f:ak} and \eq{eq:f:delta:e}. However, in certain symmetric cases the form of the Hamiltonian may be simplified. Consider, for example, the case, when the screw dislocation is directed along the $z$ axis of the Na$_3$Bi atomic lattice (or, equivalently, along the vector ${\bs K}^{(0)}$). Then, one can write the following expression for the deformations of the vielbein:
\begin{eqnarray}
\delta e_a^k &=& {\gamma_1}
K^{akj}
u^{3j} + {\gamma_2}
{\widetilde K}^{akj}
u^{3j}, \quad\quad a,k = 1,2, \quad\
\label{eNa3Bi:1}\\
\delta e_3^k &=& {\gamma_3} \epsilon_{3kj} u^{3j} + {\gamma_4} u^{3k},
\qquad\qquad\quad\ \  k = 1,2,
\label{eNa3Bi:2}\\
\delta e^3_k &=& \gamma_5 u^{3k} + {\gamma_6}  \epsilon_{3kj} u^{3j}
+
u^{3k},
\qquad\  \ \ k = 1,2,
\label{eNa3Bi:3}\\
\delta e_3^3 &=& 0, \label{eNa3Bi:4}\\
\delta e_0^k &=& 0,
\hskip 45mm  k = 1,2.
\label{eNa3Bi}
\end{eqnarray}
Here we have used the fact that the only nonzero components of the tensor of elastic deformations~\eq{eq:uij} are $u^{3i}=u^{i3}$ with $i=1,2$ given in Eq.~\eq{eq:u3a}. Moreover, we took into account that the dislocation is directed along the $z$ axis which is perpendicular to layers of honeycomb lattices formed by Na and Bi atoms in the transverse $(x,y)$ plane.
The requirement to respect the $C_3$ rotational symmetry of the honeycomb lattice in the $(x,y)$ plane allows us to define two tensors from the nearest-neighbor vectors ~\eq{eq:lll}:
\begin{eqnarray}
K^{ijk}&=&-\frac{4}{3a^3} \sum_{b=1,2,3} {l}^i_b {l}^j_b {l}^k_b
\label{eq:K}\\
\widetilde{K}^{ijk}&=&-\frac{4}{3a^3} \sum_{b=1,2,3} {l}^i_b {l}^j_b {l}^m_b \epsilon_{3mk}
\label{eq:tK}
\end{eqnarray}
which enter Eq.~\eq{eNa3Bi:1} with the material-dependent prefactors $\gamma_1$ and $\gamma_2$, respectively. The only nonzero elements of these tensors are:
\begin{eqnarray}
- K^{111} & = & K^{122} = K^{212} = K^{221} = 1\,,
\nonumber\\
\widetilde{K}^{112} & = & \widetilde{K}^{121} = \widetilde{K}^{211} = - \widetilde{K}^{222} = 1\,.
\end{eqnarray}

The tensor \eq{eq:K} was first introduced in Refs.~\cite{vozmediano3,vozmediano4}. The appearance of the second tensor structure~\eq{eq:tK} in Eq.~\eq{eNa3Bi:1} is a nontrivial fact because the tensor $\widetilde{K}^{ijk}$ is not invariant under $P$-parity transformation of the 3d space. The $P$-parity odd part is justified, however, by the chiral property of the screw dislocation, because the left-handed screws and right-handed screws are not equivalent as they cannot be superimposed on each other with the help of rotations only. Therefore, $P$-parity odd terms may appear in the Hamiltonian.

Similar arguments lead to the appearance of the other four material-dependent terms in Eqs.~\eq{eNa3Bi:2} and \eq{eNa3Bi:3} with parameters $\gamma_3, \dots, \gamma_6$. Equation~\eq{eNa3Bi} originates from the supposition that the dislocation does not break $T$ invariance so that all components of the vielbein involving one temporal and one spatial components must be zero. Notice that the deformation of the $e^0_0$ does not enter the Hamiltonian \eq{eq:HR} because $f^0_0 \equiv 1$ according to Eq.~\eq{eq:f:ak}.

One can see, that even in this relatively simple case, the expressions in Eqs. (\ref{eNa3Bi:1})-\eq{eNa3Bi} contain six phenomenological parameters $\gamma_i$, and the resulting  Hamiltonian ${\cal H}^{(R)}$, given in Eqs.~\eq{eq:H:R:transverse} and \eq{eq:H:R}, is rather complicated.

\section*{Appendix B. Certain numerical estimates}
\label{SectNUM}

We take for a reference the Dirac semimetal Cd${}_3$As${}_2$. The diffusion length of the axial charge for this semimetal was experimentally estimated in Ref.~\cite{ref:diffusion} as $L_V \approx 2 \times 10^{-6}$m. This quantity turns out to be almost temperature-independent in a wide range of temperatures $T = (50 \sim 300)$\,K.  A rough estimate of Ref.~\cite{ref:transport} gives for the relaxation time $\tau_V \sim \tau_{tr} \approx 2\times 10^{-10}$\,s. Then from Eq.~\eq{eq:L:V} one finds
\begin{equation}
D_5 = L_V^2/\tau_V \approx 2 \times 10^{-2}\mbox{m}^2/\mbox{s}\,.
\end{equation}
Correspondingly the inverse infrared cutoff $\xi_0$ may be estimated as follows:
\begin{equation}
\xi_0 \sim v_F \tau_V \sim \frac{1}{200} 300 \cdot 10^6 \, \frac{\rm m}{\rm s}\cdot 2 \times 10^{-10} {\rm s} = 3 \cdot 10^{-4}\, {\rm m}.
\label{eq:estimate:xi0}
\end{equation}
In this estimate we use the value of $v_F$ for Cd$_3$As$_2$ that is around $1/200$ speed of light.

The value of $\xi_0$ should be compared to the size of the dislocation core
\begin{equation}
\xi \sim 10^{-9}\, {\rm m}\,,
\label{eq:estimate:xi}
\end{equation}
and to the value of the diffusion length
\begin{equation}
L_V \sim 2\cdot 10^{-6}\, {\rm m}\,.
\end{equation}
Thus we see that in practice the suggested limiting case is indeed realized:
\begin{equation}
L_V \ll \xi_0
\end{equation}
and the typical value of parameter $x = \xi_0/L_V$ is $x \sim 100$.

Notice, that we used in the present paper the relativistic system of units, in which the only dimensional unit is the electron-volt (eV). For example, our distances are measured in eV$^{-1}$. We give the estimate in relativistic units, where it is expressed through eV or $1/$m, where the unit of distance (m) is related to eV$^{-1}$ according to the standard relation $[200$ MeV$]^{-1}\approx 1 $ fm $=10^{-15}$ m. Then the quantities under consideration may be expressed in the SI system using the definition of its unit of electric current (A) as ${\rm Coulomb}/{\rm s}$. The SI current equal to one Ampere corresponds to the relativistic current equal to $1/(e c)$ in the units of $1/{\rm m}$, where $e$ is the charge of electron (in Coulombs) while $c$ is the speed of light (in m$/$s).

A room temperature corresponds to $T \sim 300\, {\rm K} \approx 0.025$ eV. At the same time $D_5/c \approx 6.7 \cdot 10^{-11}\mbox{m} = 6.7 \cdot 10^4\, \mbox{fm}\approx 3 \cdot 10^{-4}\mbox{eV}^{-1} $. One should also take into account that the typical value of $v_F$ in Dirac semimetals is of the order of $\sim 1/200$ of the speed of light. We denote by $\nu$ the degree of anisotropy of the Fermi velocity. In practise in $Cd_3As_2$  \cite{semimetal_discovery2}  $v_F \hat{f}_1 \sim v_F \hat{f}_2 \sim c/200$ while $\hat{f}_3 \sim 0.1 \hat{f}_1$.  In   $Na_3Bi$ \cite{semimetal_discovery} $v_F \hat{f}_1 \approx 4.17\times 10^5 m/s$, $v_F\hat{f}_2 \approx 3.63 \times 10^5 m/s \sim c/800$,  while $v_F \hat{f}_3 \approx 1.1 \times 10^5 m/s$. Thus here $\hat{f}_3 \approx  0.27 \hat{f}_1$.

Notice, that for small values of the Burgers vector the value of $\hat{\Phi}$ may be much smaller than unity. Say, in Na$_3$Bi the minimal topological contribution to $\hat{\Phi}$ is of the order of $0.1$ (see Section~\ref{SectEMD}). However, for larger values of the component $b_3$ of the Burgers vector, the value of $\hat{\Phi}$ may always be made close to unity. Besides, the contribution of the second term to the magnetic flux in Eq.~(\ref{PhiNa3Bi}) also increases the total value of $\hat{\Phi}$. The chiral density should also be enhanced in a ``forest'' of dislocations, which are parallel to each other.

\end{document}